\documentclass[fleqn,usenatbib]{mnras}

\usepackage[T1]{fontenc}

\DeclareRobustCommand{\VAN}[3]{#2}
\let\VANthebibliography\thebibliography
\def\thebibliography{\DeclareRobustCommand{\VAN}[3]{##3}\VANthebibliography}

\usepackage{graphicx}	
\usepackage{amsmath}	
\usepackage{amssymb}	
\usepackage{newtxtext,newtxmath}
\usepackage{ulem}
\newcommand{\be}{\begin{equation}}
\newcommand{\ee}{\end{equation}}
\newcommand{\beal}{\begin{aligned}}
\newcommand{\eeal}{\end{aligned}}
\usepackage{esdiff}
\usepackage{soul}
\usepackage{xcolor}

\newcommand{\refe}{}

\usepackage{orcidlink}

\title[On the origin of the lump in circumbinary discs]
{On the origin of the lump in circumbinary discs}

\author[R. Mignon-Risse et al.]{
Raphaël Mignon-Risse\,\orcidlink{0000-0002-3072-1496},$^{1}$\thanks{E-mail: raphael.mignon-risse@apc.in2p3.fr}
Peggy Varniere\,\orcidlink{0000-0001-8888-5971},$^{2,3}$
Fabien Casse\,\orcidlink{0000-0002-8156-7628}$^{2}$
\\
$^{1}$Université Paris Cité, CNRS, CNES, Astroparticule et Cosmologie, F-75013 Paris, France\\
$^{2}$Université Paris Cité, CNRS, Astroparticule et Cosmologie, F-75013 Paris, France\\
$^{3}$Université Paris-Saclay, Université Paris Cité, CEA, CNRS, AIM, 91191, Gif-sur-Yvette, France
}

\date{Accepted XXX. Received YYY; in original form ZZZ}

\pubyear{2022}

\begin{document}
\label{firstpage}
\pagerange{\pageref{firstpage}--\pageref{lastpage}}
\maketitle

\begin{abstract}
Accreting binary black holes (BBHs) are multi-messenger sources, emitting copious electromagnetic (EM) and gravitational waves.
One of their most promising EM signatures is the lightcurve modulation caused by a strong, \refe{unique and extended} azimuthal \refe{overdensity structure orbiting} at the inner edge of the circumbinary disc (CBD), \refe{dubbed \lq lump\rq}.
In this paper, we investigate the 
\refe{origin of this structure} using 2D general-relativistic (GR) hydrodynamical simulations of a CBD in an approximate BBH spacetime.
First, we use the symmetric mass-ratio case to study the transition from the natural $m=2$ mode to $m=1$.
The asymmetry with respect to $m=2$ grows exponentially, pointing to an instability origin.
We indeed find that the CBD edge is prone to a (magneto-)hydrodynamical instability owing to the disc edge density sharpness: the Rossby Wave Instability (RWI).
The RWI criterion is naturally fullfilled at the CBD edge and we report the presence of vortices, which are typical structures of the RWI.
The RWI is also at work in the asymmetric mass-ratio cases (from $0.1$ to $0.5$).
However, the CBD edge sharpness decreases with a decreasing mass ratio, and so the lump.
By proposing a scenario for \refe{this lump} formation, our work further supports its existence in astrophysical CBDs and potential source for an EM signature of BBHs.
Finally, because the RWI is not caused by GR effects, it is also a robust candidate for the lump origin in CBDs around non-compact objects, e.g. binary protostars.
\end{abstract}




\section{Introduction}

In the attempt to understand the final stages prior to merger of binary black holes (BBHs), numerical studies have addressed the problem of circumbinary accretion flows, taking the form of a disc, around a central BBH. 
Thanks to these studies, a global picture of the accretion structures under such circumstances has emerged, including a cavity and two streams feeding the individual BHs (\citealt{macfadyen_eccentric_2008}, \citealt{shi_three-dimensional_2012}, \citealt{noble_circumbinary_2012},
\citealt{dorazio_accretion_2013},
\citealt{gold_accretion_2014},
\citealt{shi_three-dimensional_2015}, 
\citealt{ragusa_suppression_2016},
\citealt{armengol_circumbinary_2021},
\citealt{tiede_how_2021},
\citealt{liu_evolution_2021}). 
In addition to those, the most surprising feature is certainly an \refe{overdensity, localized radially and azimuthally, which is found to orbit at the Keplerian period} at the circumbinary disc inner edge\refe{, while being swept off by the spiral waves associated to the binary and therefore to the binary orbital period}.
This feature has been dubbed a \lq lump\rq\ by \cite{shi_three-dimensional_2012},\refe{ a name which refers to the aforementioned structure but does not refer to the physical phenomenon that causes it}.
The lump has the potential to produce a strong modulation of the electromagnetic flux in inclined binary systems, either directly by thermal emission or by modulating the accretion rate which then translates into a modulation in the accretion luminosity.
It is, therefore, of main interest to understand its origin, formation and growth mechanisms.

\refe{
While our work focuses on BBHs it is worth mentioning that other lump-like structures have been seen in numerical simulations across the mass domain of binary systems and also hinted at in observations of closer systems such as protoplanetary discs.
Indeed, unlike BH accretion discs, which are point-like sources (except for close-by, supermassive black holes imaged with the interferometric Event Horizon Telescope, to some extent) and make the direct observation of this lump feature impossible, protoplanetary (more specifically, transition) discs could be imaged with very high resolution thanks to the Atacama Large Millimeter/submillimeter Array (ALMA) and offered details on disc non-axisymmetries.
Among those, various observations revealed \lq horseshoe\rq\ features (\citealt{van_der_marel_major_2013}, \citealt{van_der_marel_diversity_2020}) which somehow resemble the lump feature and could possibly correspond to the same structure.
In those cases, mechanisms have been proposed (see e.g. \citealt{lyra_steady_2013}) but, as of now, none have yet prevailed. 
It has been proposed that those horseshoe features originate from a vortex trapping the dust/ice particles (\citealt{van_der_marel_vortices_2016}, \citealt{fuente_probing_2017}, \citealt{van_der_marel_major_2021}).
Some scenarii invoke the indirect influence of a planetary companion (e.g. \citealt{lyra_standing_2009}) and indeed, the accretion flows around binaries of any types share similarities.
By looking at the problem from a stronger gravity point of view we aim to see what similitudes and differences we get and how it can help us cross the gap of lump formation across the masses of binaries.
Those will be discussed in more details in Sec.~\ref{sec:disc}. 
}

 
\refe{In numerical simulations of BBHs, }lumps were found in equal-mass and in unequal-mass binaries.
Counter-intuitively, they were found to weaken with a decreasing mass ratio (\citealt{dorazio_accretion_2013}, \citealt{farris_binary_2014}, \citealt{noble_mass-ratio_2021}).
This result indicates that the nature of the lump is not associated to the native asymmetry introduced by an asymmetric BH mass distribution.

As the lump has been found to become the most prominent feature of the simulation, being a few times denser than its surroundings (\citealt{shi_three-dimensional_2012}, \citealt{noble_circumbinary_2012}, \citealt{dorazio_accretion_2013}), a phenomenological scenario of the lump feeding has been presented in \citealt{shi_three-dimensional_2012} and further in \citealt{dorazio_accretion_2013}, assuming the asymmetry has been triggered numerically (\citealt{shi_three-dimensional_2012}, \citealt{dorazio_accretion_2013}).
Both studies linked the lump growth to the disc eccentricity and to the dynamics of the streams, which consist in a runaway process increasing the asymmetry between azimuthally-opposed regions.
In this scenario, \cite{dorazio_accretion_2013} attribute the influence of the mass ratio on the lump to the torques being reduced as {the mass ratio} $q$ decreases, creating weaker shocks and smaller overdensities.
More recently, \cite{noble_mass-ratio_2021} investigated the impact of the disc magnetization: the lump is prone to form with and without magnetic fields, although, under the circumstances studied in \cite{noble_mass-ratio_2021} (a magneto-rotational unstable disc with initially poloidal magnetic fields), magnetization weakens the lump.
It is of main importance to note that, to date, the simplest physical frame to observe the lump is made of hydrodynamics with an isothermal equation of state and Newtonian gravity associated to the central binary (e.g. \citealt{macfadyen_eccentric_2008}, \citealt{tiede_how_2021}).
Without a theoretical basis for the formation of the lump, one would need to perform a numerical experiment for each set of parameters (e.g. fluid initial conditions) and physical processes in order to see if the symmetry is broken in this particular case, and to what extent.
Although lump formation appears to be a non-linear process and simulations will be needed to estimate its amplitude (and to ray-trace the system's photons to the observer to produce synthetic observables), capturing the essence of this phenomenon will allow us to somehow restrict the parameter space, in addition to providing a deeper knowledge on accreting BBHs.

In equal-mass binaries of any eccentricity, the natural symmetry of the system is the $m=2$ azimuthal symmetry.
This lump feature was, therefore, a surprise for the reason that it develops not only in unequal-mass binaries but also in the equal-mass case, and with even larger amplitude, as mentioned above.
In a recent paper focused on potential observables from the circumbinary disc around BBHs \citep{mignon-risse_electromagnetic_2022}, we - like many others (e.g. \citealt{shi_three-dimensional_2012}, \citealt{noble_circumbinary_2012}) - reported a symmetry breaking with respect to this natural $m=2$ symmetry.
There is no straighforward explanation for this behaviour.
For instance, in \citealt{gunther_circumbinary_2002}, the asymmetry was small after $90$ orbital periods and the authors attributed it to the numerical method (in particular, the matrix inversion associated to their implicit viscosity).
In this paper, we propose to use the $q=1$ case as a way to probe the lump formation, because any asymmetry becoming qualitatively significant will help us in understanding the origin of the lump.
\newline


In this work, we build on the simulations presented in the companion paper \citep{mignon-risse_electromagnetic_2022}, that we refer to for more details, in which we address the electromagnetic observables associated to pre-merger BBHs.
Those 2D inviscid general-relativistic hydrodynamical simulations were performed with the latest version of the GRMHD code {\tt GR-AMRVAC} \citep{casse_impact_2017} recently extended to any kind of spacetime metric \citep{mignon-risse_impact_2022} and based on {\tt MPI-AMRVAC} \citep{keppens_parallel_2012}.
The gravitational influence of the orbiting BBH is taken into account via an approximate analytical spacetime construction, sometimes referred to as "Near Zone" metric (see e.g. \citealt{ireland_inspiralling_2016}).
The two-dimensional grid is logathimically-stretched in the radial direction and regular in the azimuthal direction. 
{The {\tt GR-AMRVAC} code is designed to solve the general relativistic (magneto-)hydrodynamics equations translating the usual baryon density and momentum conservation laws. In the present paper we solve the hydrodynamics equations 
\be
\beal
\partial_t ({\cal D}) + \partial_j \left[  {\cal D} \left( \alpha {\rm v}^j - \beta^j \right) \right] =\ &0& ,\\
\partial_t ({\cal S}_i) + \partial_j \left(  \left[ {\cal S}_i (\alpha {\rm v}^j - \beta^j) + \alpha{\cal P}\delta_i^j \right] \right) &=& \\
-(W{\cal D}+{\cal P}(W^2-1))\partial_i \alpha + \frac{\alpha}{2} \left( {\cal S}^j {\rm v}^k + {\cal P}\gamma^{jk}\right)\partial_i \gamma_{jk} + {\cal S}_j \partial_i \beta^j &&,
\label{eq:eqs}
\eeal
\ee
where $\alpha$, $\gamma_{ij}$ and $\beta^i$ stand for the lapse function, the spatial metric elements and for the shift vector components respectively ($\sqrt{\gamma}$ stands for the square root of the spatial metric determinant). The conservative variables in a fully general framework are the relativistic density ${\cal D}$ and the relativistic momentum ${\cal S}_i$. 
$W$ stands for the Lorentz factor and ${\cal P}=\sqrt{\gamma}P$ is the pressure measured in the observer frame, with $P$ the local gas pressure. Finally, $\delta_i^j$ is Kronecker's symbol.
More details on the implementation of the code can be found in \citealt{mignon-risse_impact_2022}.
}
In this frame, we will study the evolution of a circumbinary disc orbiting the BBH, addressing the equal-mass and unequal-mass cases, focusing on the symmetry breaking and lump formation.
More information related to the numerical setup and its symmetry is given in Sec.~\ref{sec:symic}.
\newline

The paper is organized as follows.
In Sec.~\ref{sec:pbasym} we focus on the natural $m=2$ symmetry of accreting equal-mass BBHs, while its symmetry breaking is addressed in Sec.~\ref{sec:asymm}.
In Sec.~\ref{sec:rwilump}, we propose and test a (magneto-)hydrodynamical instability, the Rossby Wave Instability (RWI), as a possible origin for the growth of the asymmetry and subsequent lump formation.
After showing that the RWI is indeed at work in those simulations, we extend our scope to the unequal-mass case in Sec.~\ref{sec:unequal}.
We discuss the comparison with other works carried out in the protoplanetary context in Sec.~\ref{sec:disc}.
We present our conclusions in Sec.~\ref{sec:ccl}.

\section{Equal-mass binaries as inherently symmetric systems}
\label{sec:pbasym}
 
 In this Section, we present the physical system under study and show how it is, on the theoretical side, as well as on the numerical side, a fundamentally symmetric problem.

\subsection{{Natural symmetries and initial conditions}}
\label{sec:symic}
\begin{figure}
	\includegraphics[width=\columnwidth]{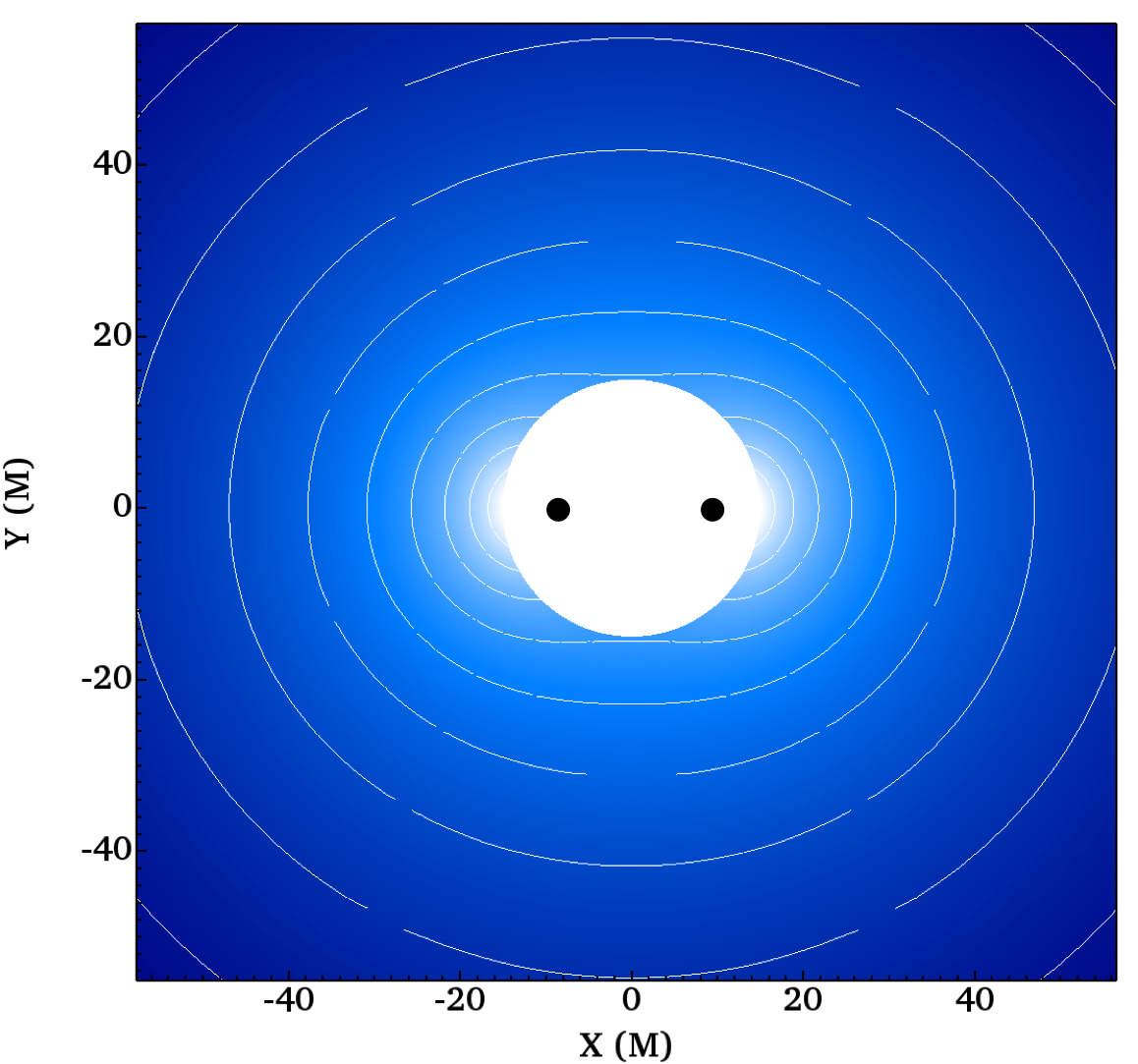}
    \caption{Map (and contours) of the radial derivative of the lapse scalar $\alpha$ from the BBH approximate metric at $t=0$. {This is the equivalent of the radial gravitational acceleration}. 
    Distances are given in units of M. 
    The $m=2$ symmetry is visible.   
    }
    \label{fig:derlapse}
\end{figure}
{When looking at an equal-mass  binary in Newtonian gravity, there is a natural symmetry in the system. Indeed, their} gravitational potential respects an $m=2$ symmetry, and this for any eccentricity.
{This stays true when} incorporating the GR effect as a background analytical metric of a binary black hole.
{This symmetry can easily be seen on} the radial derivative of the lapse scalar, {shown in} Fig.~\ref{fig:derlapse}, which is the equivalent to the radial gravitational acceleration.
{To emphasize the symmetry we overlay} the contours {on top of the color map}, with the black circles representing the {position of the} BHs.
Far from the BBH, the radial derivative of the lapse approaches spherical symmetry - it appears as axisymmetrical in the \refe{orbital} plane of the BHs.
Nevertheless, as in the Newtonian case, it transitions from spherical symmetry to azimuthal $m=2$ symmetry as the distance to the BBH decreases.
Hence, the native symmetry of the gravitational effects is an $m=2$ symmetry.

{In order to explore the origin of the late asymmetry seen in previous simulations (e.g. \citealt{macfadyen_eccentric_2008}), we need our initial conditions to preserve the natural symmetry of the system and verify that they}
do not introduce any asymmetry with respect to the $m=2$.
In a nutshell, our initial conditions are those of a disc around a single BH, they are therefore axisymmetric.
The density $\rho$ obeys a radially decreasing power-law of index $-3/4$ with an hyperbolic tangent function to set an outer disc radius around $1000\mathrm{M}$\refe{, and is equal to $1$ at about $20$~M}.
The pressure is linked to the density via a polytropic equation of state.
The radial velocity is {set to} zero and the azimuthal velocity is set so as to oppose the (GR equivalent of the) centrifugal acceleration to the gravitational acceleration diminished by the pressure gradient acceleration.
The orbital separation is kept fixed, at $b=20$~M, similar to e.g. \cite{shi_three-dimensional_2012}, \cite{noble_circumbinary_2012}, \cite{noble_mass-ratio_2021}.
Overall, these initial conditions are {fully} axisymmetric. \\

The numerical grid is axisymmetric as well.
Our runs are two-dimensional in the $(r,\phi)$ plane.
Therefore, the grid is adapted to conserving any azimuthal symmetry within the code's accuracy.
The grid covers radii $[15,1500] \mathrm{M}$ (for the equal-mass run) and azimuthal angles $[0,2\pi]$ with $784\times400$ cells. Gradient-null boundary conditions are set in the radial direction. Moreover, no material is allowed to enter the grid at the outer radial boundary.
\newline

Overall, the physical system under study and its numerical translation {at $t=0$} respect an $m=2$ symmetry.

\subsection{{Early stage and preserved symmetry}}

\begin{figure}
	\includegraphics[width=\columnwidth]{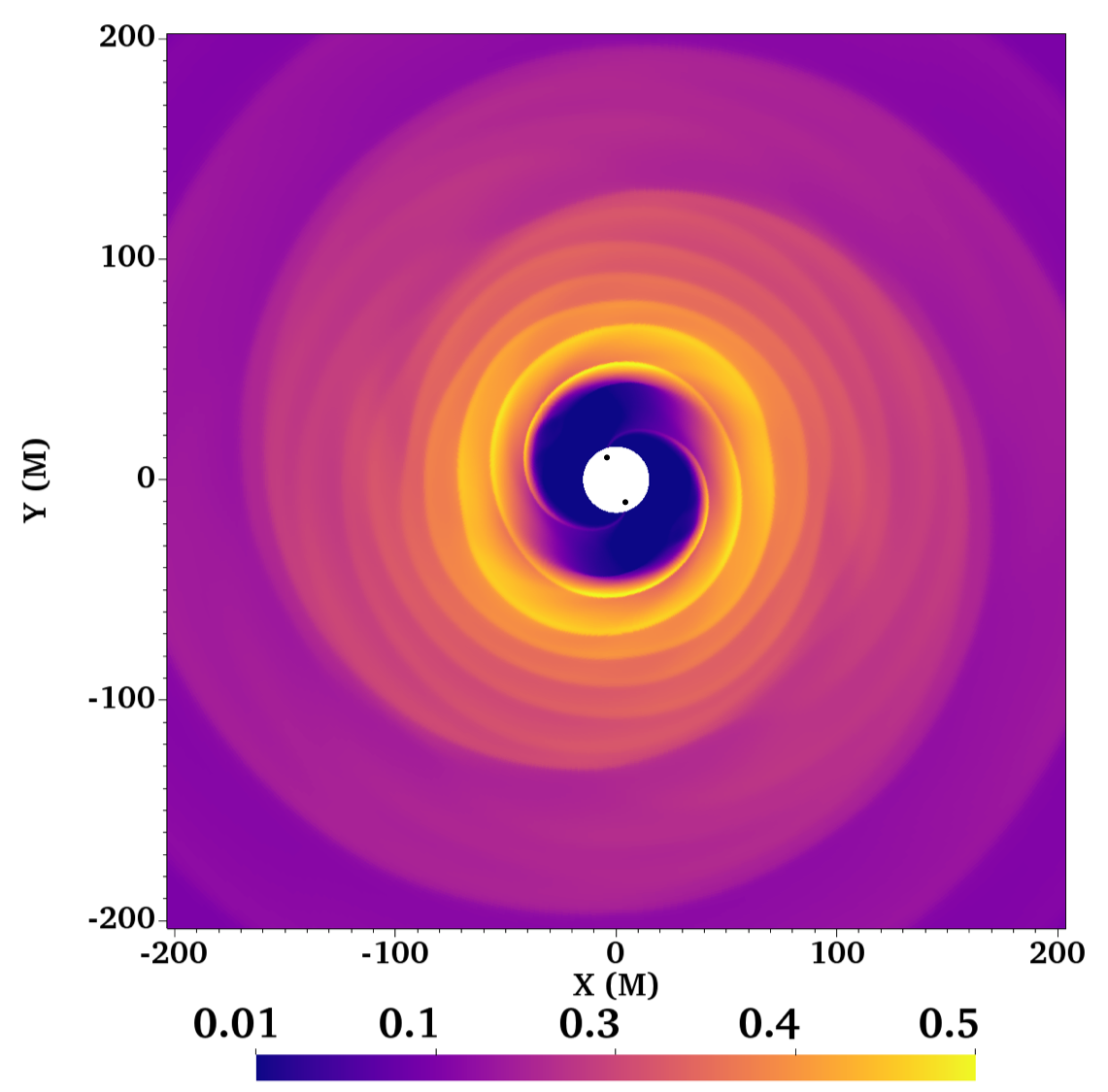}
    \caption{Density map at
    $t=24000\mathrm{M}$ (${\approx}40~\mathrm{P_{orb}}$) in the $q=1$ run, when the system is qualitatively symmetric.
    Distances are given in units of M. 
    The BBH is located within the inner boundary of the grid. }
    \label{fig:rhomap_q1}
\end{figure}

\begin{figure}
	\includegraphics[width=\columnwidth] {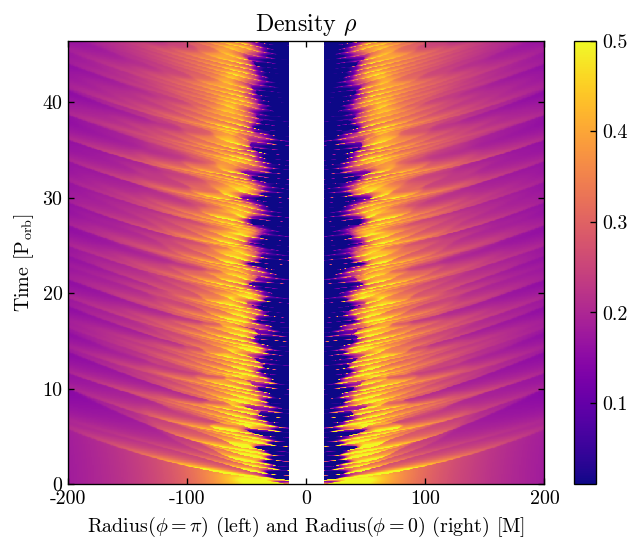}
	\includegraphics[width=\columnwidth] {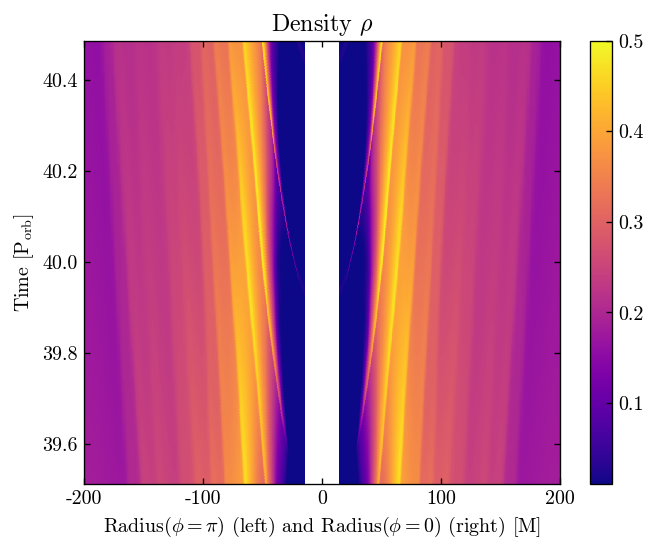}
    \caption{Radius-time map of the density (linear scale, code units) in the $q=1$ run\refe{ from $t{\approx}0$ to $t{\approx}45\, \mathrm{P_{orb}}$ (top panel) and temporal zoom around $t{\sim}68\, \mathrm{P_{orb}}$ (bottom panel)}.
    Distances are given in units of $\mathrm{M}$. 
    Two-arm spiral waves appear as diagonal lines of enhanced density launched from the binary\refe{ and connect to the binary as streams}.
    The two sides are qualitatively similar.}
    \label{fig:r_t_q1_symm}
\end{figure}

{This symmetry is actually preserved in the early stage of the simulation as shown on Fig.~\ref{fig:rhomap_q1}, which depict the density map (and two black holes) after about $40$ orbits 
of the binary. For clarity we only show a zoom in the inner $400$~M of the simulation. }
{Another way to visualize the symmetry is presented in} Fig.~\ref{fig:r_t_q1_symm} which shows the {evolution of the} density for 
{a 1D cut through the center of mass} as a function of the radius and time {(here taken for $\phi=0$ and $\phi=\pi$ to get both sides of the disc)}.
This last visualization is useful to observe any periodic feature, the propagation of waves and{, in our case,} the eventual loss of symmetry as a function of time.

First of all, the inner region below $r{\approx}2\mathrm{b}$ {is briskly} cleared of material, forming a low-density cavity, as reported {previously} in the literature (e.g. \citealt{artymowicz_dynamics_1994} \citealt{macfadyen_eccentric_2008}, \citealt{hirsh_cavity_2020}).
From $t=0$ to ${\sim}50~\mathrm{P_{orb}}$, the density obtained along the $\phi=0$ and $\phi=\pi$ directions is qualitatively similar, indicating that if asymmetries are present, they are negligible (e.g. \citealt{dorazio_accretion_2013}).
Outgoing spiral waves appear as overdense diagonal lines travelling away from the BBH.
Accretion onto the BBH occurs via two identical streams.
Those appear as the filaments connecting the circumbinary disc to the innermost boundary of the grid in Fig.~\ref{fig:r_t_q1_symm}.\\

To sum up, we showed that the equal-mass BBH is inherently symmetric, from its theoretical model to the numerical experiment meant to study it.
For about $50~\mathrm{P_{orb}}$, the system preserves its native azimuthal $m=2$ symmetry.


\section{Breaking the symmetry: from $m=2$ to $m=1$}
\label{sec:asymm}

{While we have seen that the $m=2$ symmetry of the system is kept for more than $40$ orbits of the binary, the previous works cited above have shown that \refe{one} lump or \refe{one azimuthal overdensity (attributed to an $m=1$ structure, see \citealt{shi_three-dimensional_2012})
}
appears in the later stage of  similar simulations. Here we aim to study the rise and growth of the} deviation from the natural $m=2$ symmetry in our equal-mass binary simulation {in the hope
to garner clues on its origin.}

\subsection{Deviation from the $m=2$ symmetry}

{While Fig.~\ref{fig:r_t_q1_symm} showed that the density was qualitatively similar, as we are trying to find how that symmetry gets broken in later stages of the simulation
we need a more quantitative way to look at the deviation. 
In order to do that we follow the evolution of the} maximal 
{deviation from the $m=2$ symmetry as a function of time}.
To do so, we first compute the maximal value of the density for all azimuthal angles, denoted $\max(\rho(\phi))$.
For each angle $\phi$ between $0$ and $\pi$, we compute the difference $\max(\rho(\phi))-\max(\rho(\phi+\pi))$, which should be identically zero if the $m=2$ symmetry of the system was perfectly conserved.
Finally, we take the maximal value of this quantity with respect to $\phi$ {to collapse the array to one value that can be followed as function of time}.
This allows us to identify the largest deviation from $m=2$ symmetry {in each snapshot}.\\

\begin{figure}
	\includegraphics[width=\columnwidth]{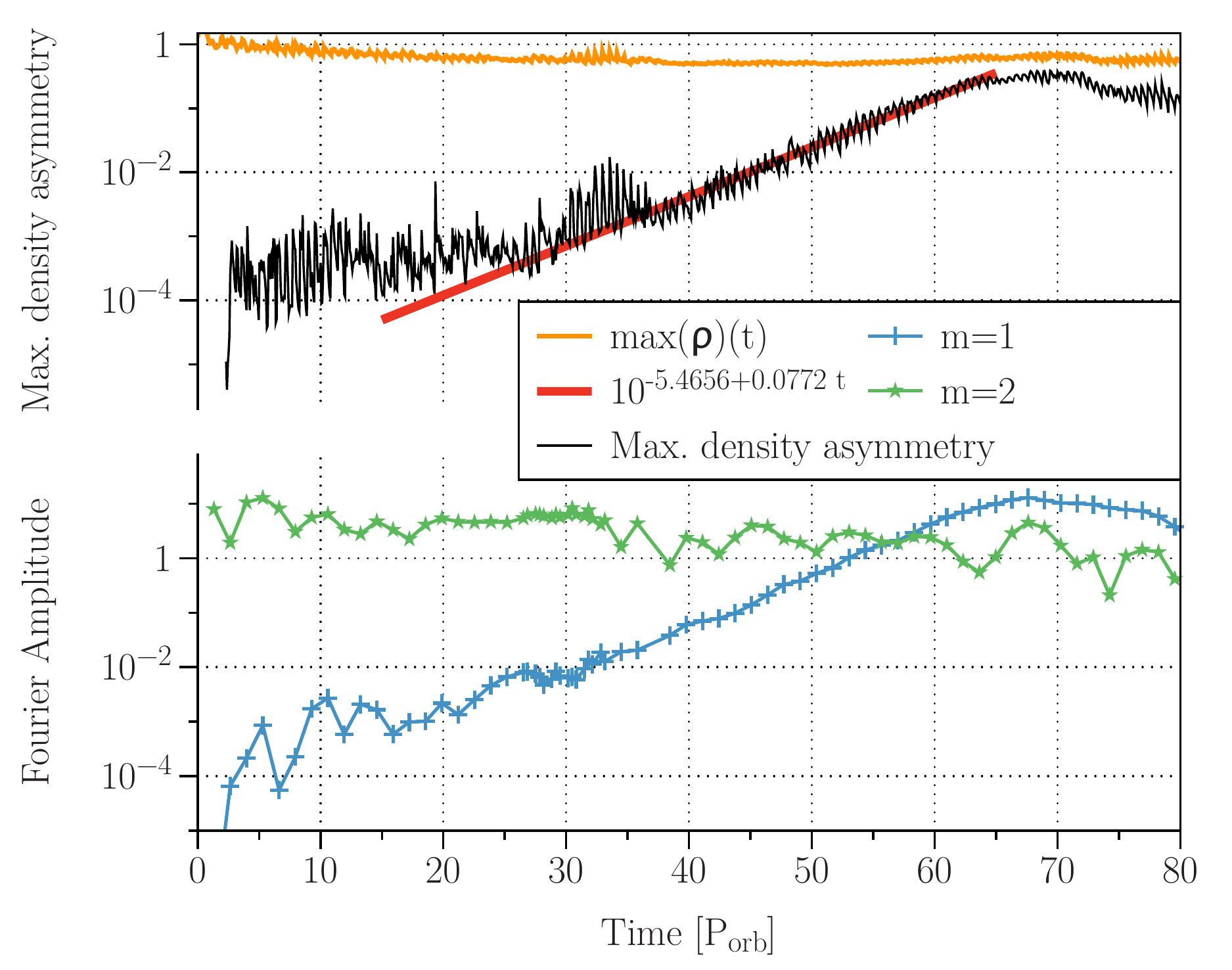}
    \caption{
    \refe{Top panel: }measure of the deviation with respect to the natural $m=2$ symmetry ($q=1$) in the density distribution\refe{ computed as $\max(\max(\rho(\phi))-\max(\rho(\phi+\pi)))$.
    This quantity is referred to as the \lq maximal density asymmetry\rq\ in the main text}.
    The red line is a fit by an exponential function.
    One might expect the $m=2$ natural symmetry of the system to be conserved, hence the plotted quantity would be of the order of the code precision. 
    For comparison, the maximal density in the simulation is shown in orange as a function of time.
    \refe{Bottom panel: evolution of the two modes dominating at the end of the simulation: $m=2$ in green, $m=1$ in blue.}}
    \label{fig:growth}
\end{figure}

\refe{Top panel of }Fig.~\ref{fig:growth} shows the {evolution of the} aforementioned quantity, that we will {now} refer to as \lq maximal density asymmetry\rq, as a function of time.
{We see that for the first} $2$ orbits, the density distribution follows the $m=2$ symmetry {at a precision} below $10^{-5}$ {which could be
related to the numerical precision of the metric used here (\citealt{mignon-risse_electromagnetic_2022})}. 
{Then, for} more than $20$ orbital periods, the maximal density asymmetry {stays} of the order of $10^{-3}${, hence its lack of detectability on the snapshots of the early phase of the simulation.}
{But,} after about $30$~orbits the maximal density asymmetry {starts to increase exponentially}.
{Indeed,}  the curve {can be fitted between} ${\approx}30$ to ${\approx}60$ orbital periods, {in  semi-log} by a function of type $10^{\mathrm{a}t+\mathrm{b'}}$ 
shown by the red line.
Hence, {this proves that, starting from a small and nearly constant value,} the maximal density asymmetry grows exponentially with time in our simulation.
After roughly $60$ orbital periods, it saturates around $0.1-0.3$.
For comparison, the inner edge of the circumbinary disc before symmetry breaking has densities of $0.5-1$.
Thus, after growing exponentially, the asymmetry we report becomes quantitatively important and the system now strongly deviates from its natural $m=2$ symmetry.

\refe{Alternatively to this view, we perform a Fourier analysis and plot the evolution of the modes that eventually dominate at the end of the simulation in the bottom panel of Fig.~\ref{fig:growth}.
We perform the Fourier analysis on the region located between radii $15$~M (the innermost boundary of the simulation box) and $175$~M.
While all cells are not at the exact same time in such a GR simulation, we checked that the lapse derivative (which controls the time dilation) in the region of the circumbinary disc edge - where the lump appears - is weak enough to explore the possibility of a Fourier analysis (e.g. \citealt{casse_impact_2017}).}

\refe{The nearly constant forcing of the $m=2$ mode, visible in the bottom panel of Fig.~\ref{fig:growth}, comes from the binary. Meanwhile, the amplitude of the $m=1$ increases exponentially, in agreement with the previous study on the maximal density asymmetry shown in the top panel of the same Figure.
Eventually, the $m=1$ mode becomes the dominant one and saturates after roughly $60$
orbital periods, coexisting with the $m=2$.}
\\

\subsection{The $m=1$ asymmetry, or lump}

{Similarly to previous studies, our simulation shows}
that a density asymmetry with respect to the $m=2$ symmetry becomes significant after a few tens orbits, { while also quantifying its growth and when it starts}.
{We can now focus on the 2D picture} 
of this asymmetry{, namely} its features and how the system evolves as it appears. \\

\begin{figure}
	\includegraphics[width=\columnwidth] {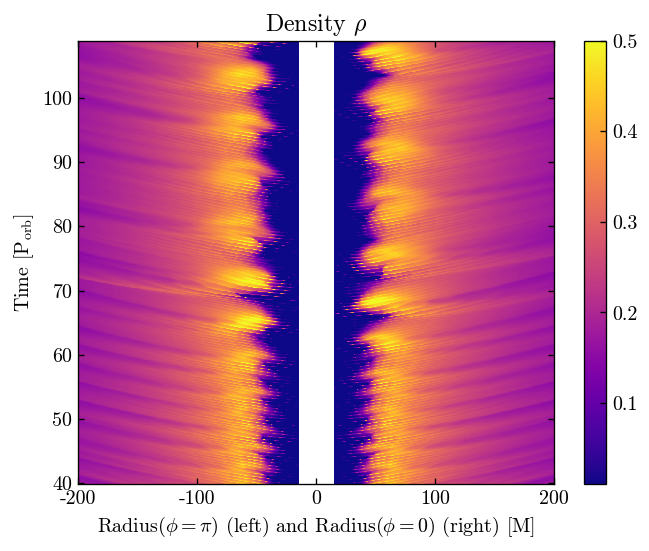}
	\includegraphics[width=\columnwidth] {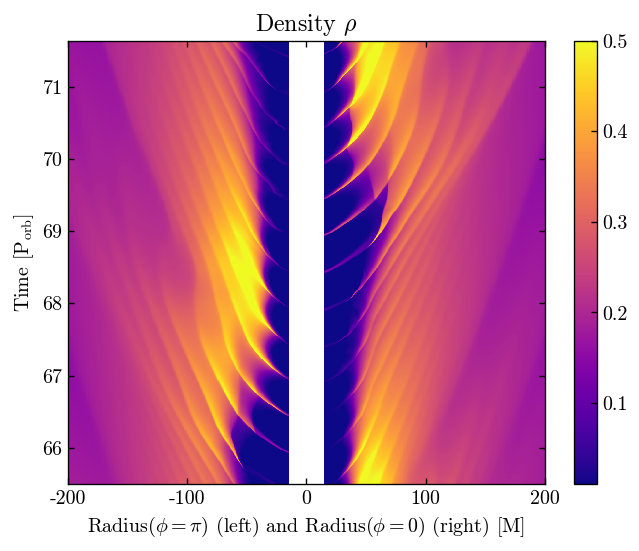}
    \caption{\refe{Top panel: r}adius-time map of the density (linear scale) in the $q=1$ run\refe{, from $40\, \mathrm{P_{orb}}$ to $110\, \mathrm{P_{orb}}$}. 
    The $m=2$ symmetry breaking becomes visible after $t{\approx}57\mathrm{P_{orb}}$.
    Then, the lump appears as the $m=1$ dense spot{, first visible on one side then on the other as it \refe{orbits} around the circumbinary disc}.
    The period associated with the lump is longer than the period of the spiral waves, i.e. longer than the semi-orbital period.
    \refe{Bottom panel: temporal zoom over roughly one lump's period and that the $m=2$ spiral waves are deviated from their original trajectory (they do not follow diagonal lines) at the lump location in the $(r,t)-$plane, as compared to Fig.~\ref{fig:r_t_q1_symm}.}}
    \label{fig:r_t_q1}
\end{figure}

\begin{figure}
	\includegraphics[width=\columnwidth]{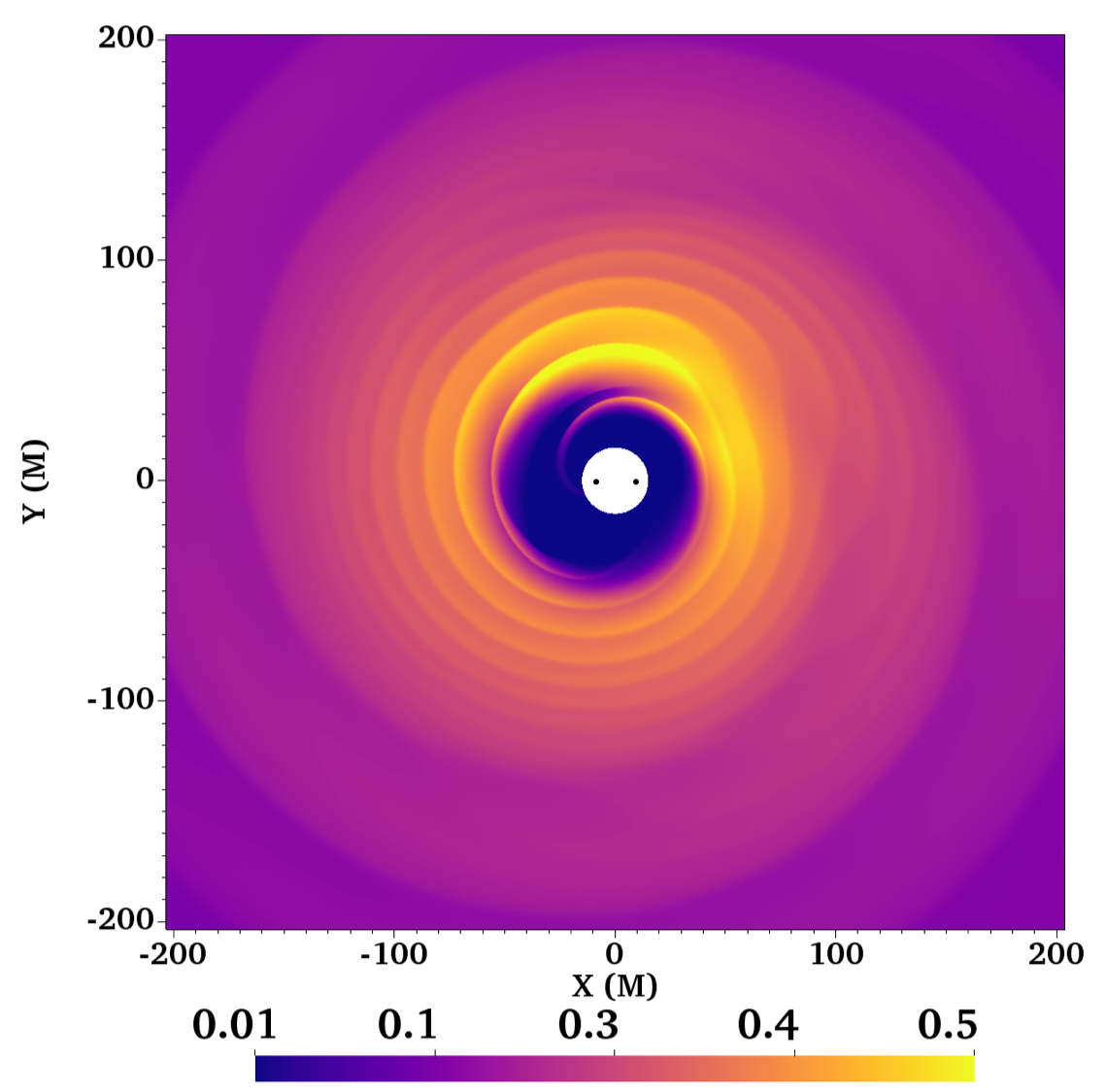}
    \caption{Density map at $t=38000\mathrm{M/c}$ (${\approx}63~\mathrm{P_{orb}}$) in the $q=1$ run.
    The BBH is located within the inner boundary of the grid.
    The lump is visible as the densest region in the bottom panel.
    This structure is reminiscent of the behaviour of the RWI (see Fig.~2 of \citealt{varniere_living_2020}).}
    \label{fig:rhomap_q1_lump}
\end{figure}

{In order to show how the growing deviation from the $m=2$ symmetry translates visually we plot in
\refe{the top panel of }Figure~\ref{fig:r_t_q1} the}
continuation of Fig.~\ref{fig:r_t_q1_symm} in the stage where the asymmetry becomes visible.
Around $60~\mathrm{P_{orb}}$. when the maximal density asymmetry reaches about $0.1$ (i.e. $25\%$ of the maximal density in the simulation at that time), we start to see that the filaments are not in sync for both phases but alternate,
\refe{ as expected from }an odd azimuthal mode \refe{(see bottom panel of Fig.~\ref{fig:growth})}.
{This is even clearer on} Fig.~\ref{fig:rhomap_q1_lump}{, which} shows {the inner regions} of the density map in the asymmetric epoch.
{Indeed, we can see that the so-called lump \citep{shi_three-dimensional_2012} is visually consistent with} an azimuthal \refe{overdensity} {on top of the earlier spiral structure seen of Fig.~\ref{fig:rhomap_q1}}.
\refe{It is also shown, in the bottom panel of Fig.~\ref{fig:r_t_q1}, that the $m=2$ spiral waves are deviated from their original trajectory (they do not follow 
diagonal lines) around the lump location in the $(r,t)-$plane, as compared to Fig.~\ref{fig:r_t_q1_symm}.}  
\\

Knowing that the lump structure is an $m=1$, we can use Fig.~\ref{fig:r_t_q1} to see the periodicity at which the lump is orbiting. We see that it is larger than the (semi-orbital) 
period of the spiral waves seen from the start of the simulation. 
Moreover, outgoing spiral waves associated with the lump feature are visible: they are emitted at the lump location and their periodicity appears to be the same as the lump's.
{Hence confirming the presence of two structures in the disc, the original spiral with a $m=2$ symmetry and the lump. This is especially seen as the stream closer to the lump is
stronger than the one at the opposite phase.}\\
The qualitative picture depicted above - cavity, streams, lump - is consistent with the numerous numerical studies of circumbinary discs as presented in the introduction.
Nevertheless, one thing that was not mentionned previously is how the $m=1$ structure of Fig.\ref{fig:rhomap_q1_lump} is reminiscent of {the $m=1$ structure seen at the inner edge of single black hole discs in} simulations focused on the {development of the} Rossby Wave Instability (RWI, see for example Fig.~2 of \citealt{varniere_living_2020}).
{This} motivated us to assess whether the RWI could be responsible for the lump.

\section{The RWI as a possible origin for the lump}
\label{sec:rwilump}

{While there is a long history of detecting this lump in circumbinary discs, there was little progress in uncovering the physical mechanism giving it rise.
Following on the visual similitude between the density map in Fig.\ref{fig:rhomap_q1_lump} and previous work on the Rossby Wave Instability}
we investigate whether it could be responsible for breaking this symmetry and causing the formation of the lump.\\

The RWI has been {found to occur} in various astrophysical {systems}, from galactic discs \citep{lovelace_negative_1978} to accretion discs around black holes and protoplanetary discs
 (e.g. \citealt{varniere_reviving_2006}); for a review, we refer the reader to \cite{lovelace_rossby_2014}. {One of the needed characteristics in common to a lot of those systems
 is the existence of a sharp disc edge, such as the one occuring naturally} 
at the innermost stable circular orbit of discs around single BHs.
Since the circumbinary disc {also} possesses a sharp inner edge, {it seems natural to look if} the RWI could develop there {as well, especially as}
the behaviour of the {inner region of the} circumbinary disc, shown in Fig.~\ref{fig:rhomap_q1_lump} is in qualitative agreement with the presence of the RWI.

{Indeed,} we reported in Sec.~\ref{sec:pbasym} a symmetry breaking with respect to the $m=2$ mode in equal-mass binaries and the further exponential growth of the maximal density asymmetry{,
leading to a strong $m=1$ mode}.
Such an exponential growth is consistent with the presence of an instability.
Therefore, in this Section, we address the possibility that the RWI is developing in this system by first looking at its criterion, then by searching for vortices as they are a distinct feature of this instability.

\subsection{Basics of the RWI}
\label{sec:rwi}

{The un-magnetized RWI is characterized by the} 
presence of a local extremum of the  vortensity, which is defined as (in the non-relativistic case)
\be
\mathcal{L} = \frac{\nabla \times \mathbf{v}}{\Sigma},
\label{eq:vort}
\ee
where $\mathbf{v}$ is the fluid velocity vector and $\Sigma$ is the surface density. 
The above {criteria} is for a  2D, vertically-integrated density, view {but it can be extended to magnetized discs\footnote{{In the presence of magnetic field, we need an extremum of the 
magnetized vortensity defined as $\mathcal{L_B} = (\nabla \times \mathbf{v}) \Sigma/B^2$.}  }\citep{li_rossby_2000,tagger_accretionejection_2006}}
{as well as for full} 3D simulations \citep{meheut_rossby_2010}, in Newtonian gravity and GR (\citealt{casse_rossby_2018} with BH spin).
{This is important in trying to explain the formation of the lump by the RWI as the lump has} been reported in unmagnetized (e.g. \citealt{dorazio_accretion_2013}) and magnetized (e.g. \citealt{shi_three-dimensional_2012}) circumbinary disc simulations, in {both} 2D (e.g. \citealt{macfadyen_eccentric_2008}) and 3D (e.g. \citealt{noble_circumbinary_2012}) {simulations and threfore the instability needs to exist 
under all those conditions}.\\

The development of the RWI is associated with large-scale spiral density waves and Rossby waves emitted away from the vortensity extremum region, 
{inside which a Rossby vortex will form}.
{As long as the criteria is fulfilled, the}  instability will grow exponentially with time {and gas will concentrate in the vortex region. Such}
accretion onto the Rossby vortices has been invoked as a possible mechanism for planet formation within protoplanetary discs {and could explain here the creation of the massive lump}.
{While the early dominant mode of the RWI depends on local conditions and excitation (\citealt{casse_impact_2017}), it tends to cascade toward lower $m$ modes and often the $m=1$.
Hence,} the development of the RWI {will eventually} break the $m=2$ axisymmetry. \\

{While it was first difficult to find} extrema in vorticity, {it was shown that they are} naturally present at the ISCO of black hole accretion discs due to GR effects (e.g. \citealt{vincent_flux_2013}), 
or in  a density bump (see \citealt{falanga_general_2007} for an application to the flares of Sagittarius A*) making those discs locally unstable to the RWI.
Here, we show that an extremum of vortensity can {also} naturally appear at the inner edge of the circumbinary disc.

\subsection{Is the instability criterion fullfilled ?}

\begin{figure}
	\includegraphics[width=\columnwidth]{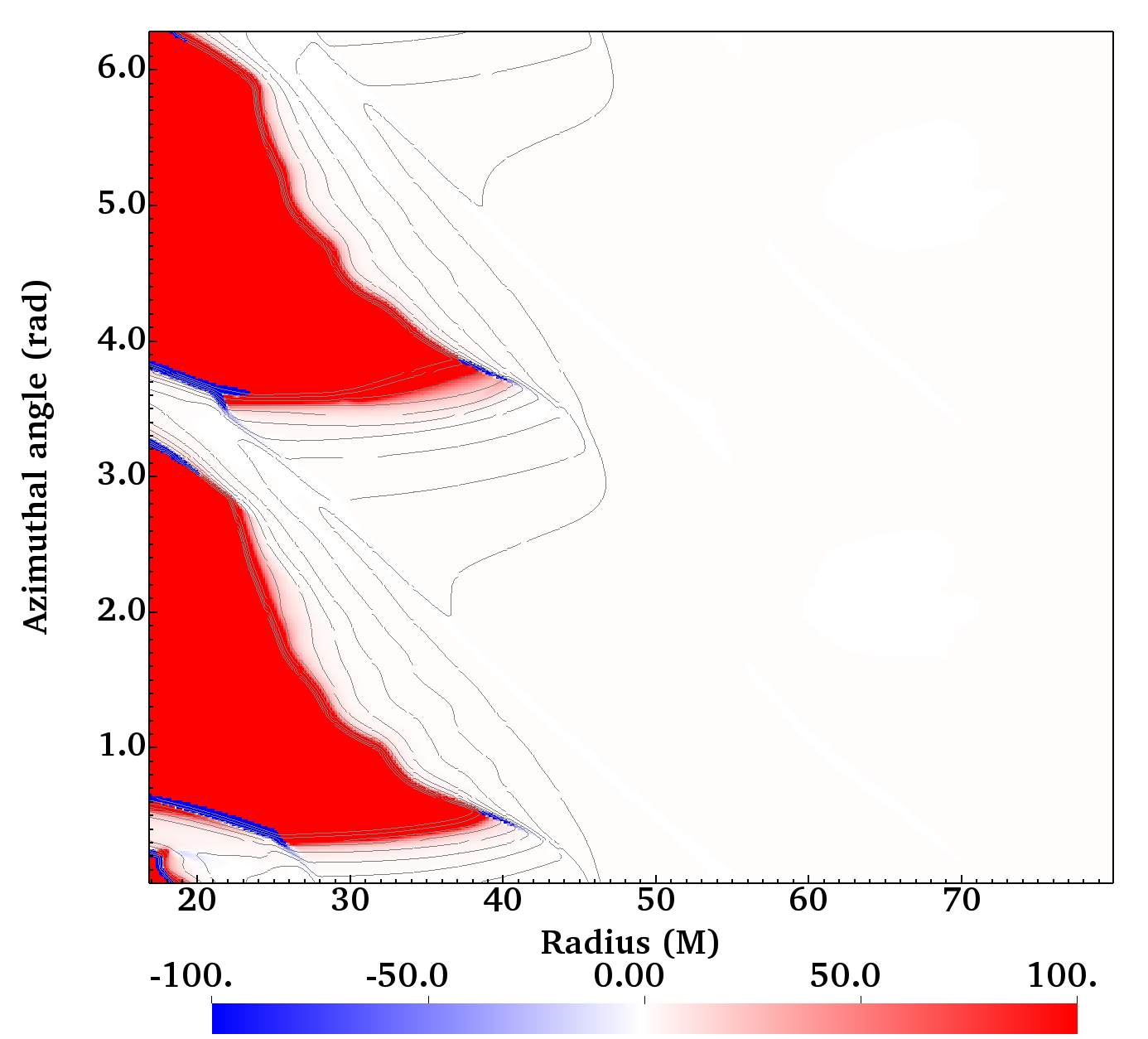}
	\includegraphics[width=\columnwidth]{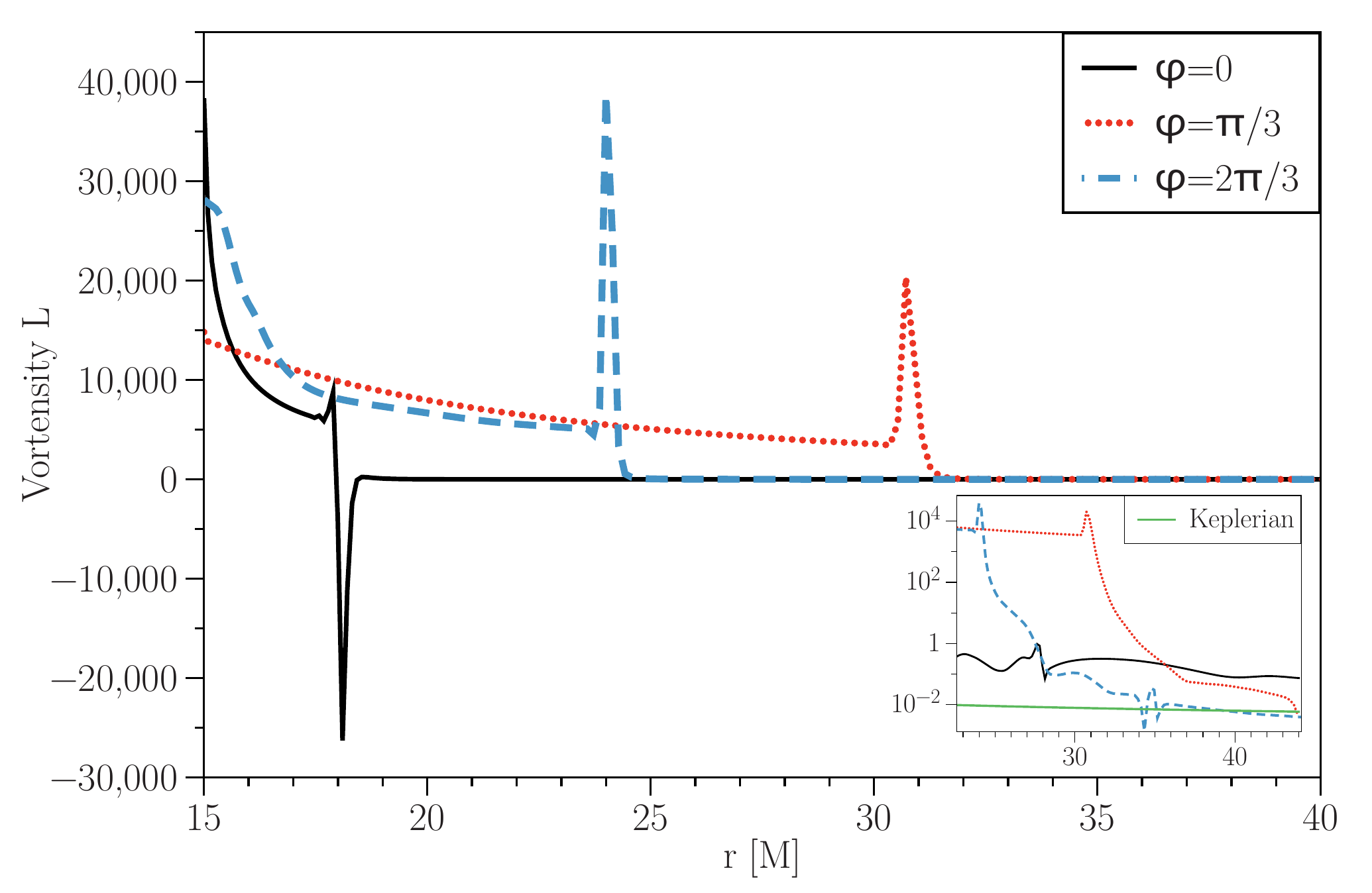}	
    \caption{
    Top panel: $(r,\phi)$-map of the vortensity at time $32000\mathrm{M}$ (${\approx}53\mathrm{P_{orb}}$) for $q=1$, with the radius in units of $\mathrm{M}$. The contours of $\log(\rho)$ are overplotted in gray.
    The red-colored region corresponds to the cavity.
    Bottom panel: radial profile of the vortensity \refe{in linear scale and in logarithmic scale in the zoom-in view}
    for three azimuthal angles: $\phi=0,\pi/3,2\pi/3$ drawn from the top panel plot. Extrema of the vortensity are visible at all azimuthal angles but various radii, following the inner edge of the circumbinary disc. It is, therefore, unstable to the Rossby wave instability. The $\phi=0$ vortensity profile crosses one of the two streams (see the top panel).
    \refe{For reference, we plot the Keplerian profile of the vortensity around a single object of mass $\mathrm{M}$ in green in the zoom-in view.}}
    \label{fig:vort}
\end{figure}

In order to investigate whether the RWI {indeed} develops in our simulation, we first {need to show that the} instability criterion is fullfilled.
As  no GR equivalent {to the Newtonian criteria presented in Eq.~\ref{eq:vort}} has been derived yet, {we use a similar approach as in \cite{casse_rossby_2018} and} 
compute the vortensity using the velocity $\tilde{v}^2_\phi = v^\phi  v^ig_{\phi i} \approx (v^\phi)^2\gamma_{\phi \phi}$ {since $\gamma_{\phi\phi}$ is the dominant component of the spatial metric in the $\phi$ direction at the location of the lump}.\\

{Contrary to previous work on the RWI, we have a fully 2D system and a radial cut does not give the full representation of the criteria so we first show on}
Fig.~\ref{fig:vort} the vortensity in the $(r,\phi)$ plane (top panel, with the contours of $\log(\rho)$ overplotted) and {the more typical 1D radial plot, but for} 
various azimuthal angles (bottom panel), at $t=32000\mathrm{M}$ (${\approx}53\mathrm{P_{orb}}$).
We plotted the vortensity for azimuthal angles between $0$ and $\pi$ because, at this stage, the system is still close to the $m=2$ symmetry, as visible on the top panel ($5-10\%$).
{Both of those representations of the vortensity} show that the edge of the circumbinary disc, i.e. the outermost radius of the cavity{ shown as the red-colored region in the top panel}, is a region {of high} vortensity {variation over a small radial extent}.
The bottom panel {examplifies}, for three {azimuths}: $\phi=0$, $\phi=\pi/3$ and $\phi=2\pi/3$, {how the} extremum in vortensity {behaves along the edge of the cavity}. 
{It is worth noting the different behaviour} when crossing a stream, {indeed} the vortensity both meets a positive and a negative local extremum. 
This is visible at $\phi=0$ on the bottom panel of Fig.~\ref{fig:vort}.\\

We showed that the RWI instability criterion is fullfilled in the $q=1$ simulation {for} nearly all azimuthal angles {along the edge of the cavity/circumbinary disc, hence 
giving strength to the notion that the RWI is responsible for the long-lived lump that appears after the stabilization of the cavity}.
Moreover, {as} the cavity is {continually being carved by the presence of the binary, even as accretion is trying to fill it,} the inner edge of the circumbinary disc remains sharp and prone to the RWI.

\subsection{{Can we detect vortices at the circumbinary edge?}}
\label{sec:vortices_q1}

\begin{figure}
	\includegraphics[width=\columnwidth]{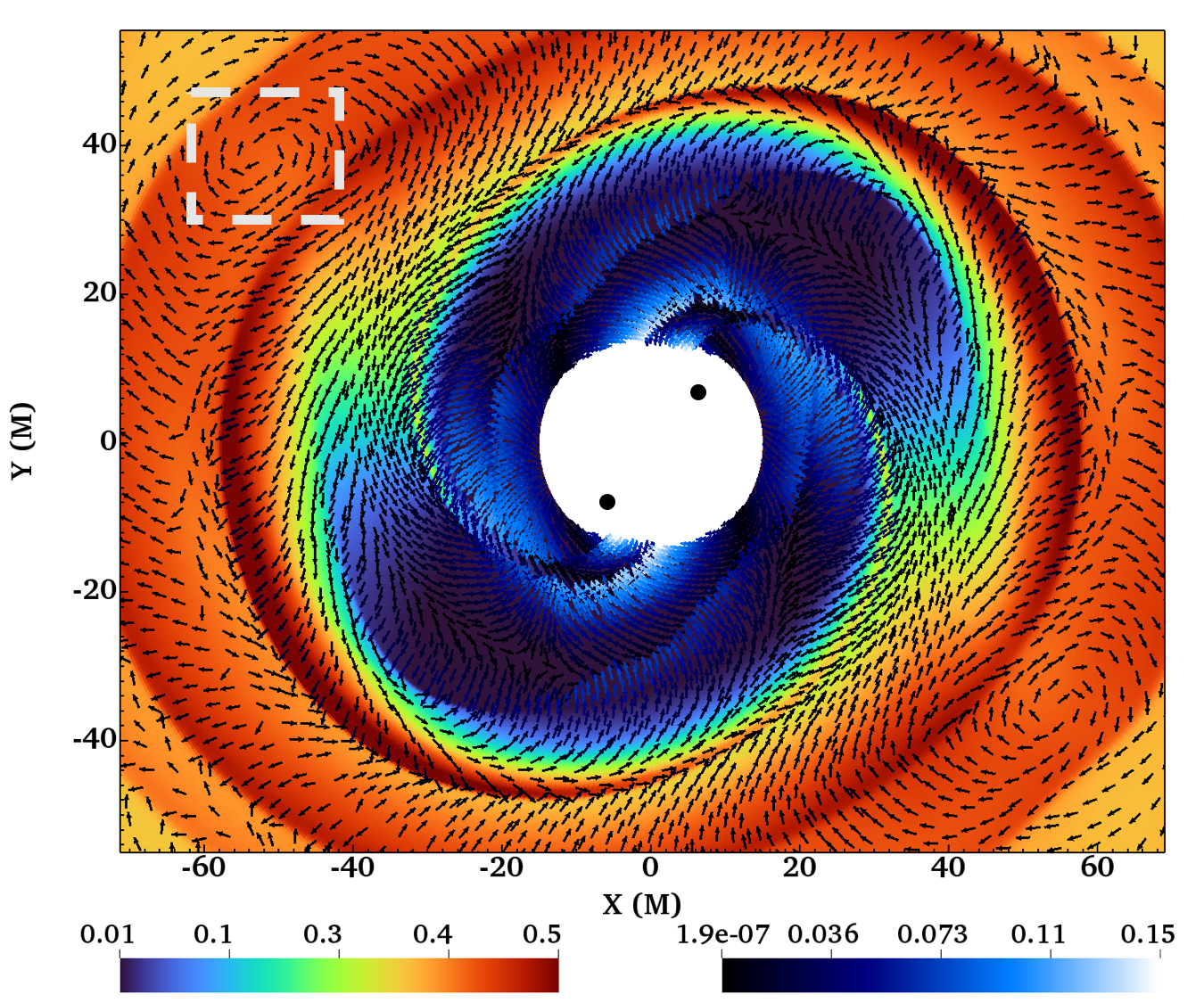}
	\includegraphics[width=\columnwidth]{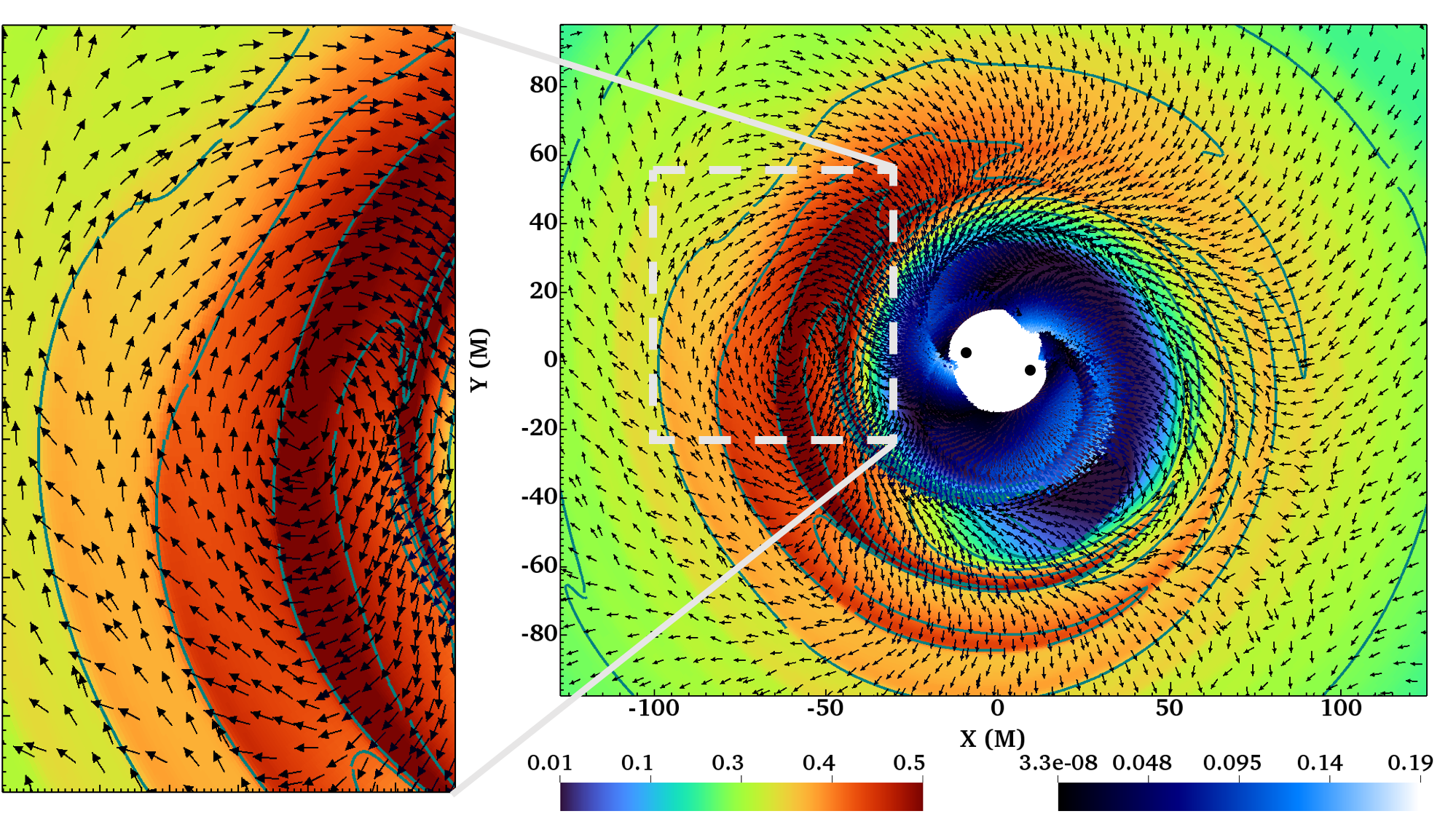}
    \caption{\refe{Top panel: }The vector field $v^i(t)-v^i(t=0)$ is overlayed on top of the density map at $t{\approx}20000\mathrm{M/c}$ (${\approx}33\mathrm{P_{orb}}$) in the high-resolution version of run $q=1$.
    The left colorbar refers to the density and the right colorbar to the velocity, both in code units.
    Distances are in units of $M$.
    It exhibits two vortices symmetric with respect to the BBH ; one is indicated by the lightgray box.
    The presence of {long-lived} vortices is typical of the RWI \citep{lovelace_rossby_2014} {and displayed by very few other instabilities}.
    \refe{Bottom panel: focus on the lump and vortex region at $t{\approx}39200\mathrm{M/c}$ (${\approx}65\mathrm{P_{orb}}$), i.e. during the asymmetric epoch when the m=1 mode reached saturation. Density contours are overlayed in blue to show how the density distribution is affected by the RWI vortex.
    Let us note that, because of the nearly constant $m=2$ forcing by the binary sweeping through the vortex, the density maximum is not necessarily located in the vortex.}}
    \label{fig:vortex_q1}
\end{figure}
{The most unique characteristic of the RWI is the formation of long-lived} 
vortices. 
{Hence, after showing that the criteria of the instability is fullfilled, finding those vortices near the edge of the cavity/circumbinary disc 
would make the RWI a very likely candidate for the creation of the lump.}\\

When looking at a disc with a large azimuthal velocity it is difficult to clearly see vortices as each vortex is a perturbation of the overall large 
velocity field. In order to overcome that we choose to plot the  $v^i(t)-v^i(t=0)$ vectors instead, emphasizing the vortices' presence\refe{ even in early times when the $m=1$ mode is weaker than the $m=2$, hence emphasizing that it is the same phenomenon which also occurs at later times}.
Figure~\ref{fig:vortex_q1} shows the density map {of the inner cavity and circumbinary disc} with the velocity field $v^i(t)-v^i(t=~0)$ vectors {overlayed to show the vortices'
presence near the edge}.
For visibility {and easier identification of the vortices}, we visualize the {velocity field on Fig.~\ref{fig:vortex_q1} with} a slightly higher-resolution run than exposed otherwise. 
\refe{In the top panel, t}wo azimuthally-opposed vortices are visible, at locations $[-50,40]$ (indicated by the lightgray box, for visualization purposes) and $[50,-40]$, {which is} right beyond the {maximal radial extent of the precessing} circumbinary disc edge.
They are symmetric because the snapshot is taken {at ${\approx}33\mathrm{P_{orb}}$, which is still in the early stage of the RWI growth, and before the merging of the vortices in a dominant $m=1$ 
mode\refe{ as displayed in the bottom panel.}
Moreover, they are not just velocity features: the density field around those is increasingly affected with time\refe{, as illustrated in the zoom-in view of the bottom panel, where the density contour is overlayed in blue}.
The presence of vortices strongly suggests the development of the RWI in our simulation {with the merging of the vortices into a $m=1$ dominant mode.
\newline

\begin{figure}
	\includegraphics[width=\columnwidth]{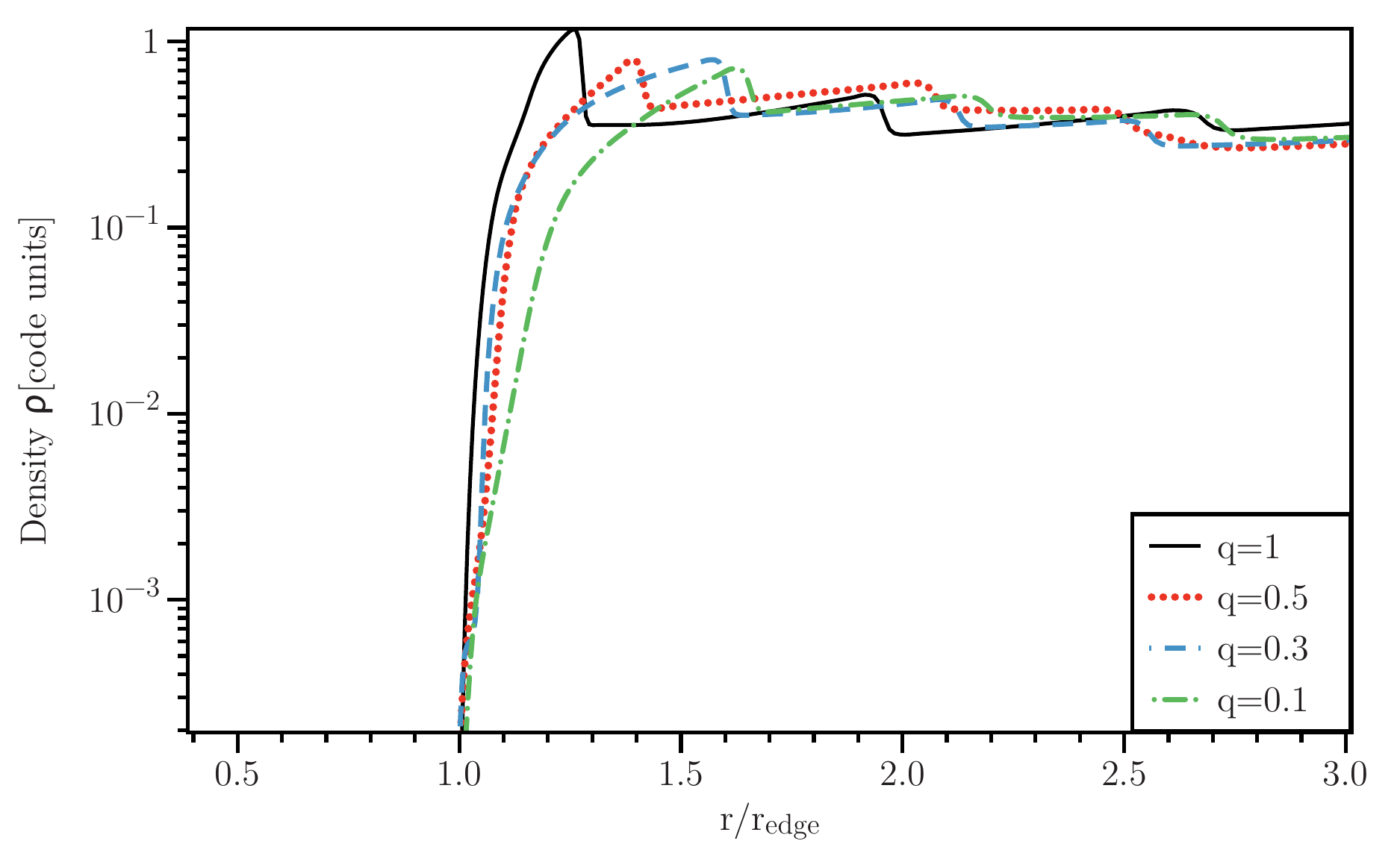}
	\caption{{Sharpness of the} radial profile of the density as a function of the mass ratio, for $q=[1,0.5,0.3,0.1]$ in black, red, blue, and green, respectively.}
    \label{fig:r_rho_q}
\end{figure}

\begin{figure}
	\includegraphics[width=0.95\columnwidth]{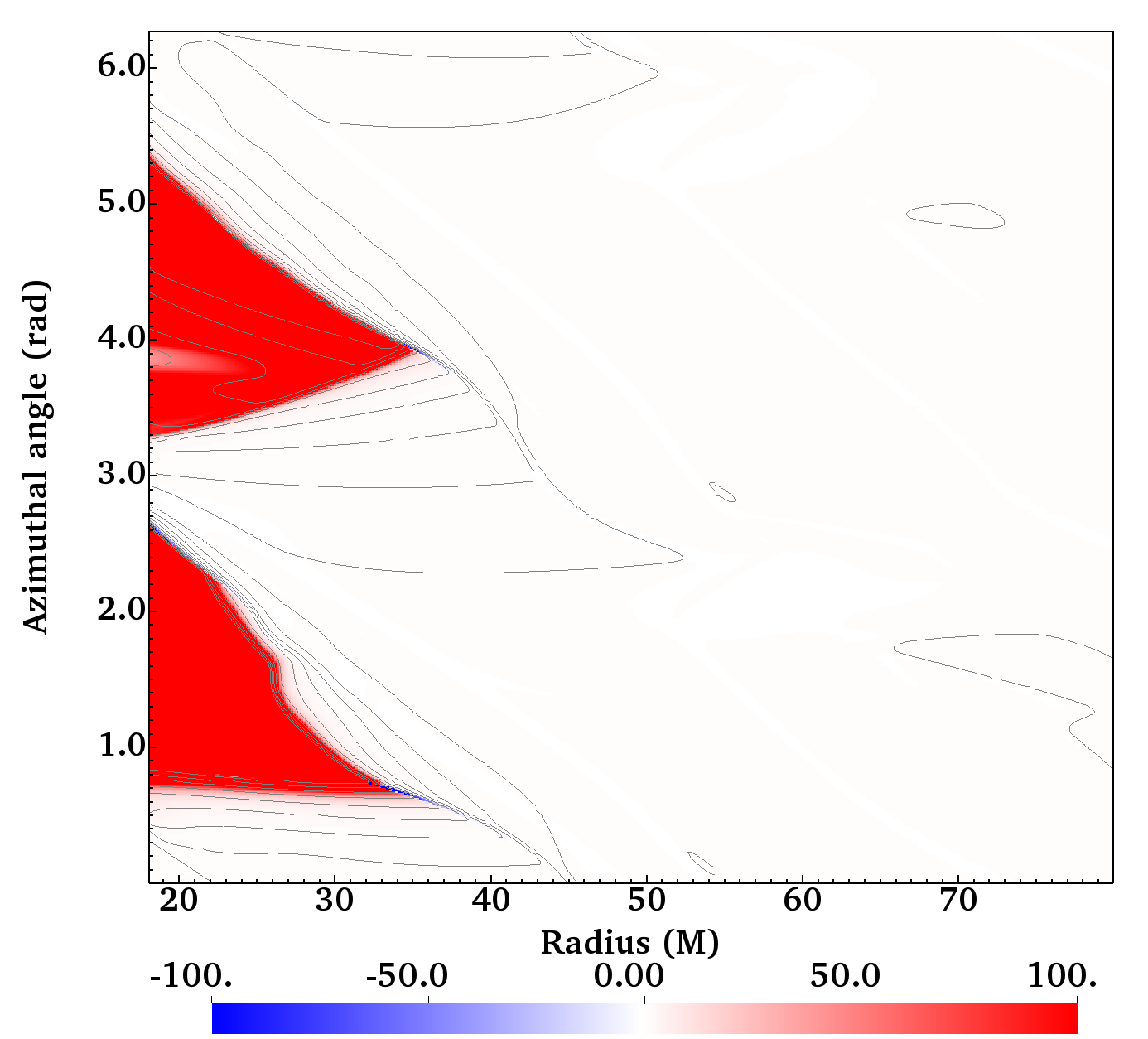}
	\includegraphics[width=0.95\columnwidth]{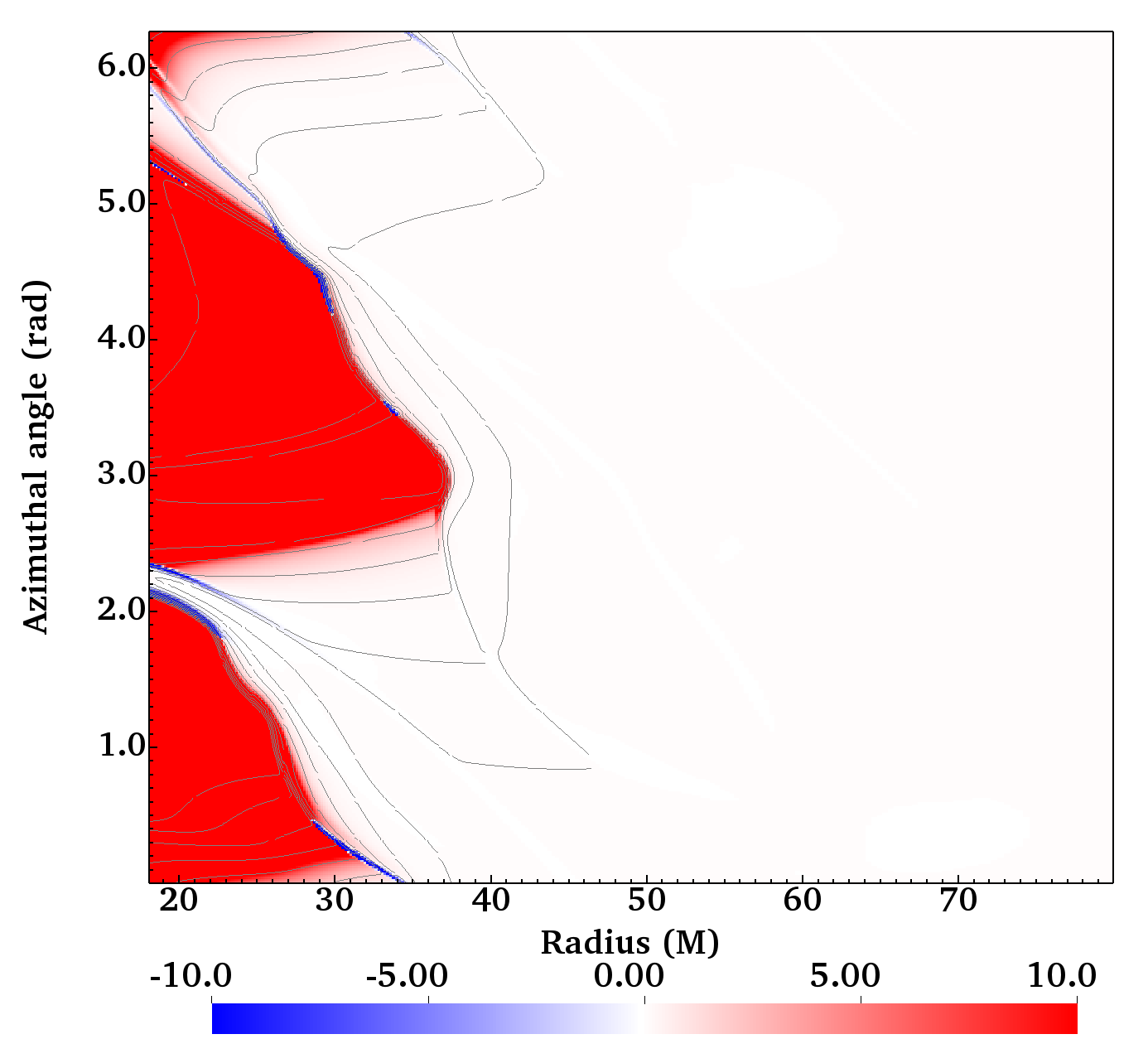}
	\includegraphics[width=0.95\columnwidth]{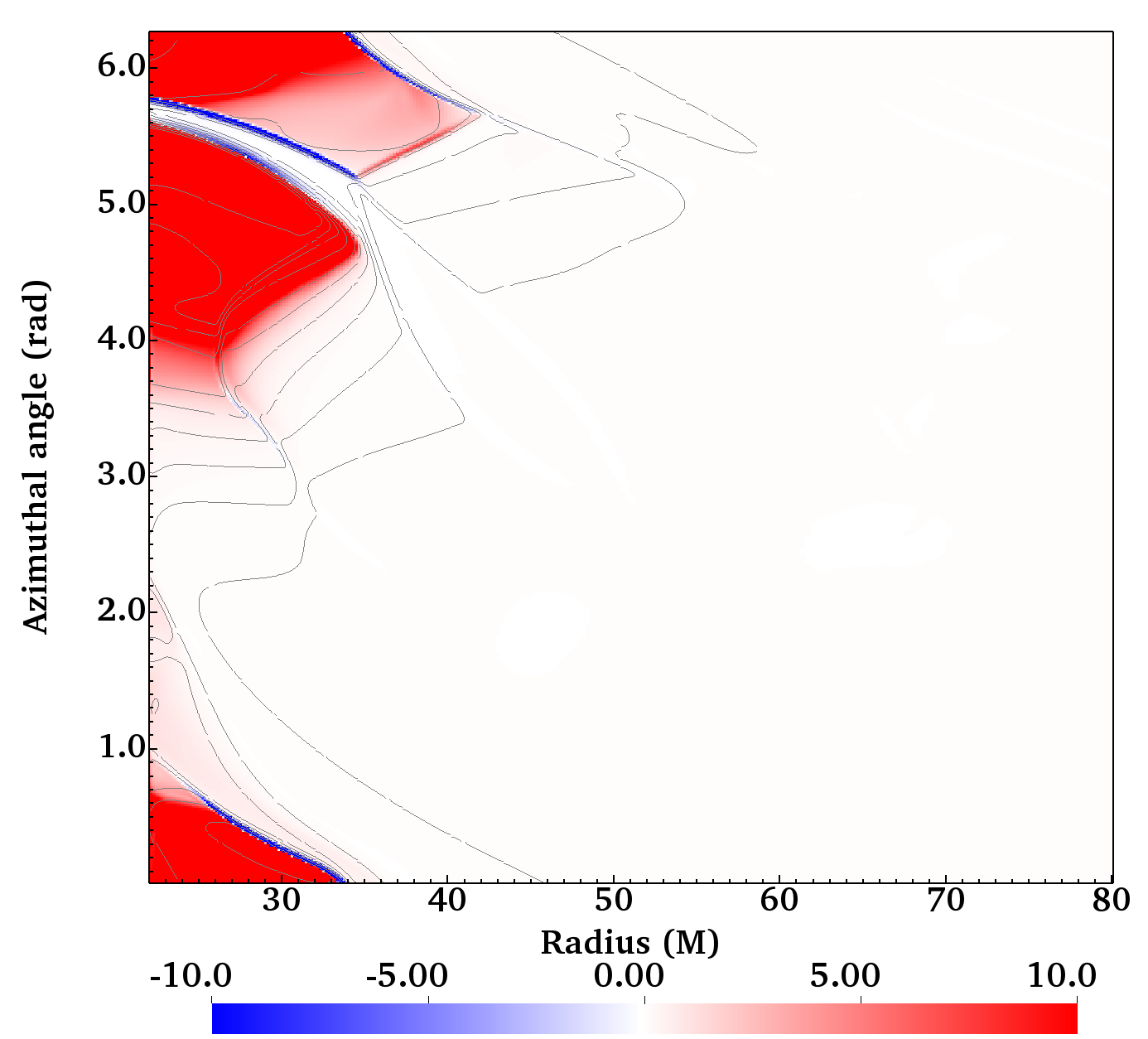}
 \caption{From top to bottom: map in $(r,\phi)$ of the vortensity for $q=[0.5,0.3,0.1]$, with the radius in units of $\mathrm{M}$. The contours of $\log(\rho)$ are overplotted in gray. 
    The red-colored region corresponds to the cavity. }
    \label{fig:vort_qne1}
\end{figure}

{Thanks to the presence of vortices along the edge of the circumbinary disc, and their subsequent merging into a strong $m=1$ mode, we propose the RWI as}
a model for {the formation and continual feeding of the} lump
in circumbinary discs.
\refe{The exact density maximum is not necessarily found at the location of the vortex because of the region is being continuously swept off by the $m=2$ spiral waves excited by the binary.
However, the vortex indeed orbits at the local Keplerian period, as was already reported for the lump in the literature and in contrast with the $m=2$, whose period is the binary orbital period.}
In the $q=1$ case, the seed for the instability is numerical and likely associated with the BBH approximate metric in the present case, for which any operation involving the metric deals with numbers close to machine precision.
Such a numerical seed has been invoked as well in the literature to explain the asymmetry of the streams {as well as} has been proposed to be at the origin of a runaway process resulting in the 
lump formation (\citealt{shi_three-dimensional_2012}, \citealt{dorazio_accretion_2013}).
Nevertheless, we have followed the development of the RWI in our simulation from a slight asymmetry to the binary $m=2$ to a strong $m=1$ vortex accumulating matter similarly to what is 
seen at the edge of planet gaps or dead zones in protoplanetary discs
\citep{varniere_reviving_2006}\refe{, with the difference that the density maximum is not necessarily located in the vortex here because of the strong density fluctuations carried by the spiral waves sweeping through it.
Nonetheless, the major point is that the overdensity dubbed lump needs the $m=1$ to exist, even though it is not always at the same location because of the spiral waves.}

\section{Extension to asymmetric mass ratios}
\label{sec:unequal}

{Up to now we have focused on the} symmetric case $q=1$ {where there is an unexpected} transition from {the system} native $m=2$ symmetry to an $m=1$ mode.
{We used this break of symmetry to help us identify its origin as} the development of the RWI {at the edge of the circumbinary disc}.
{The $q\neq 1$ systems lack this initial symmetry and we therefore} expect the natural asymmetry of the system to help seeding the instability.
Hence, we check for the presence of the RWI {for a few} $q\neq 1$ cases.
To do so, we perform a similar study as in Sec.~\ref{sec:rwilump}, namely we address the presence of vortensity extrema as diagnostics of a RWI-unstable state.
{Once we saw that the criteria was indeed fulfilled in those cases}, we looked for the presence of vortices close to the circumbinary disc inner edge.\\

In the following, we consider three mass ratios: $q=0.5$, $q=0.3$ and $q=0.1$ to explore the impact of the initial asymmetry on the existence of RWI and its ability to create
\refe{an azimuthal overdensity consistent with the lump.}
As we have seen earlier, the RWI tends to evolve near sharp edges, so as the first direct impact of the mass ratio on the growth of the RWI we look at the evolution of the radial profile of the density in the perpendicular direction to the cavity edge in order to probe the {compared} circumbinary disc inner edge sharpness.
Figure~\ref{fig:r_rho_q} shows
the profiles  for $q=[1,0.5,0.3,0.1]$ {with the radius for each simulation} normalized to the radius of their disc edge {so that we can compare the sharpness of the edge}.
{We see that the $q=1$ case has the sharpest edge, namely the strongest gradient, and hence is more favorable to the RWI.
Hence, while we expect the RWI
to also appear for all the $q$ values presented here, the growth rate of the instability will likely decrease with $q$ and there might be a value of $q$ small enough that the 
edge of the circumbinary disc is not sharp enough to launch the RWI. 
The value of $q$ where this happens will depend strongly on the equation of state and magnetic field strength. 
Looking for this value is beyond the scope of this paper, which is aiming at understanding the origin of the lump.}
\newline

{Thanks to these density profiles, we have seen that not all mass ratios might be able to fullfill the criteria of the RWI. 
We now look at it in more detail and plot the} counterpart of Fig.~\ref{fig:vort} for $q\neq 1$.
Figure~\ref{fig:vort_qne1} shows the vortensity maps in the $q=[0.5,0.3,0.1]$ runs (from top to bottom) {with} the contours of $\log(\rho)$ {in overlay}. 
Because of the asymmetric mass ratio, the two opposed cavities are different as they form, in even less than one orbital timescale, and plotting the vortensity at the symmetric epoch as for $q=1$ is not possible.
The red regions correspond to the cavity, which {look} increasingly asymmetric as $q$ decreases from $0.5$ to $0.1${, nevertheless,} a vortensity radial extremum is present at the 
location of the circumbinary disc inner edge. 
For conciseness, we do not display the radial profile of the vortensity as in the bottom panel of Fig.~\ref{fig:vort}{, especially as} similar peaks as in 
the $q=1$ case are obtained.
{One major difference is that} they do not appear for every value of the azimuthal angle $\phi$, unlike the $q=1$ case.
Indeed, as shown in Fig.~\ref{fig:vort_qne1}, a density radial extremum (one way to produce a vortensity extremum) is not present at all azimuthal angles, especially for $q=0.3$ and $q=0.1$. 
Since the RWI criterion relies on the vortensity extremum, our results indicate that {while} the circumbinary disc edge is RWI-unstable {the azimuthal size of this} 
unstable region {decreases with} $q$, {starting from} $2\pi$ in the $q=1$ case.
A weakening of the lump with a decreasing mass ratio was actually observed in the literature (\citealt{dorazio_accretion_2013}, \citealt{noble_mass-ratio_2021}).
The present work suggests that, indeed, as a result from the reduced azimuthal range of unstable regions to the RWI, the growth rate is reduced as the mass ratio decreases.
\newline

\begin{figure}
	\includegraphics[width=\columnwidth]{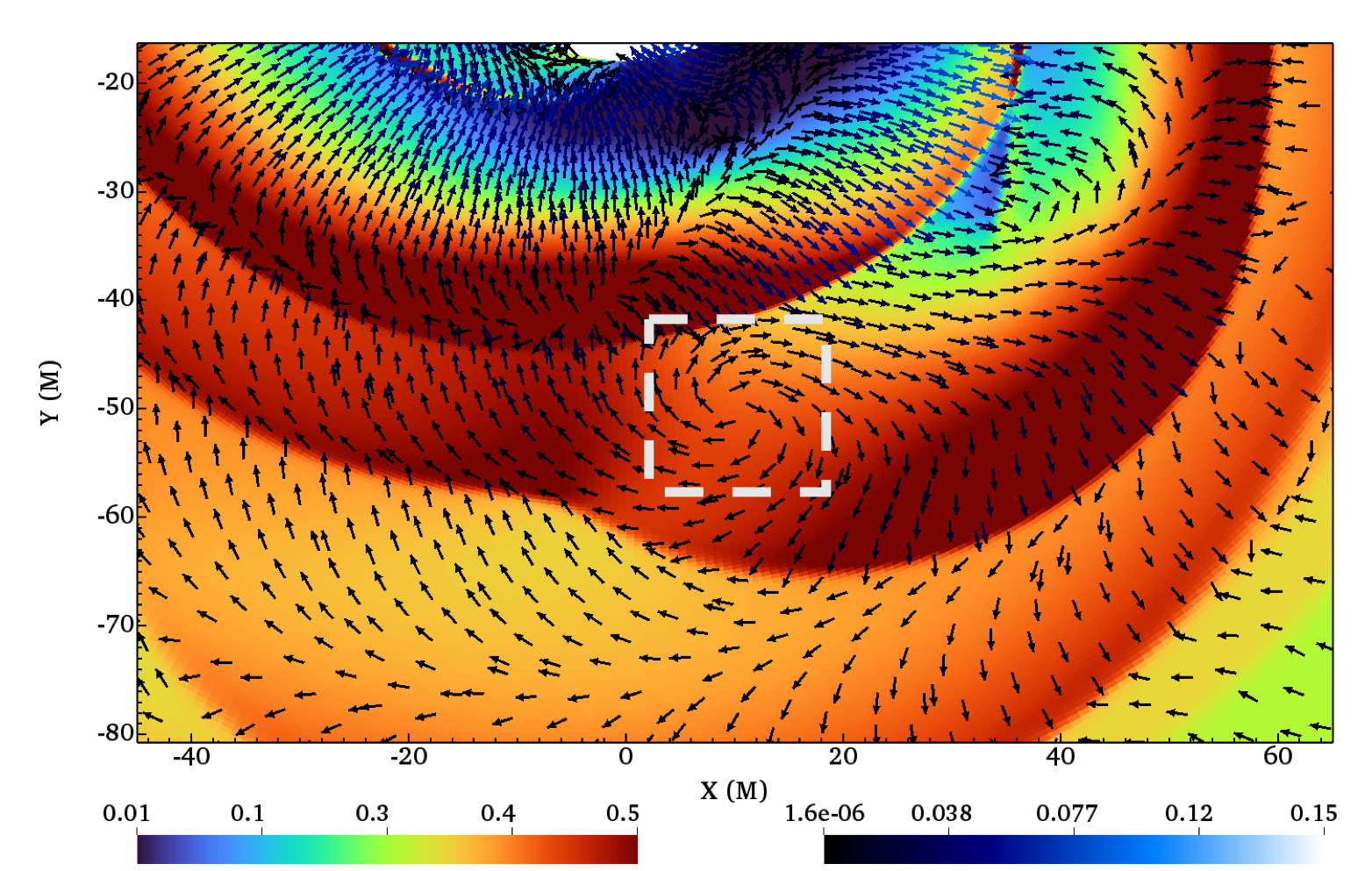}
	\includegraphics[width=\columnwidth]{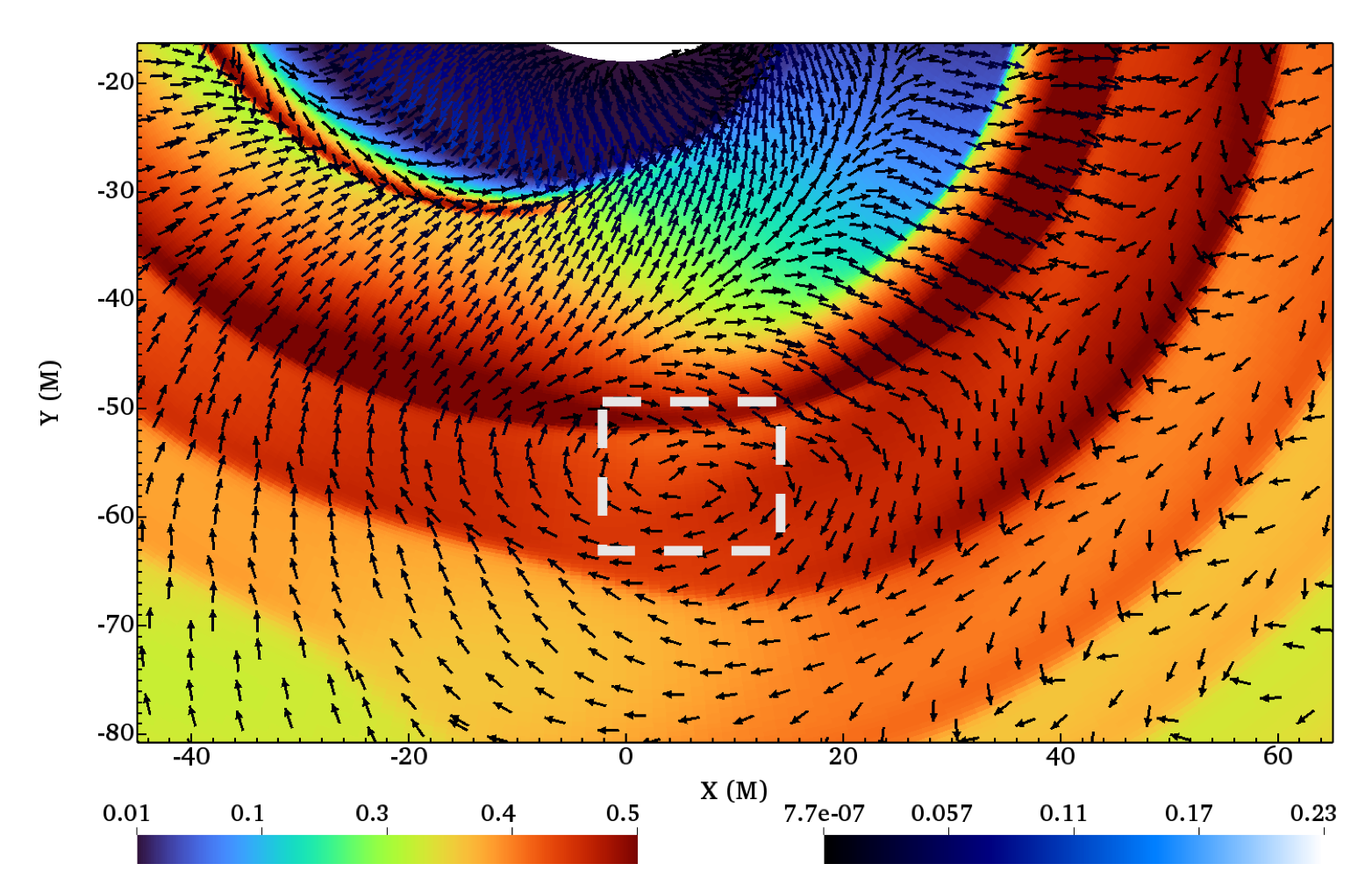}	
	\includegraphics[width=\columnwidth]{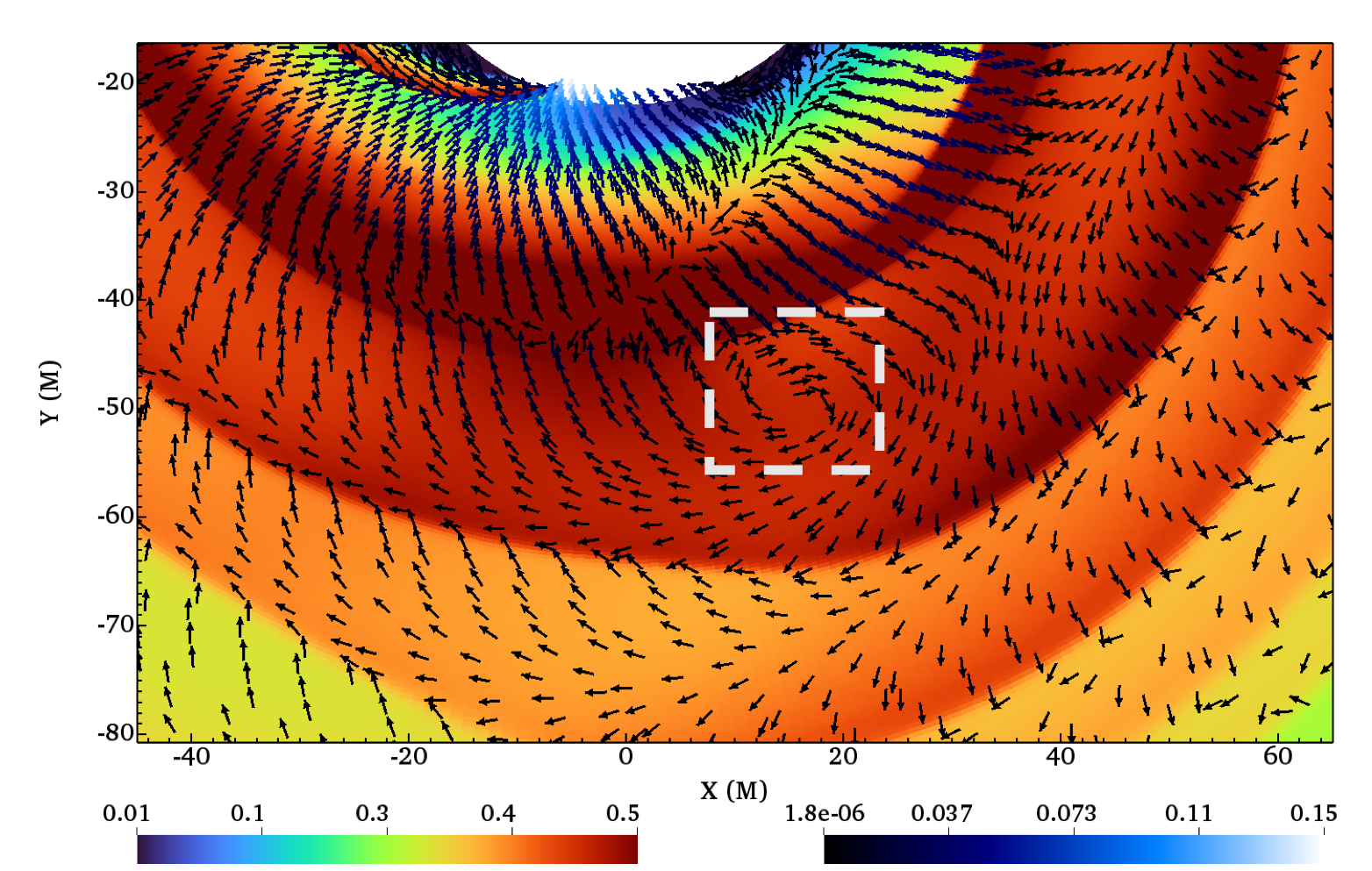}
	\caption{{Similar to Fig.~\ref{fig:vortex_q1} for different mass ratios.} From top to bottom: density maps for $q=0.5$ ($t=4000 \mathrm{M}{\approx}7\mathrm{P_{orb}}$), $q=0.3$ ($t=12000 \mathrm{M}{\approx}20\mathrm{P_{orb}}$), $q=0.1$ 
	($t=4000 \mathrm{M}{\approx}7\mathrm{P_{orb}}$).
	The velocity vectors are overplotted. 
	Vortices, indicated by lightgrey boxes, are observed for every value of $q$ from $0.1$ to $1$.
	The left colorbar refers to the density and the right colorbar to the velocity, both in code units.}
    \label{fig:vortex_qne1}
\end{figure}

{As a last} diagnostic {of} the development of the RWI, we look for vortices in the {different} $q\neq1$ runs {with} Fig.~\ref{fig:vortex_qne1}.
This Figure is similar to Fig.~\ref{fig:vortex_q1} but for $q=[0.5,0.3,0.1]$ (from top to bottom), {with the major difference being much earlier in time, meaning} after {just} 
a few orbits {instead of a few tens of orbits}.
{We see that indeed,} vortices become visible roughly after the cavity has been cleared, for any {of the} $q$ values {that we simulated}.
This Figure also illustrates how the density distribution is visibly affected in the region where vortices are found (this is also the case for $q=1$ but less visible on Fig.~\ref{fig:vortex_q1}).
\\

{We have shown that, for $q$ between $1$ and $0.1$, the edge of the circumbinary disc is unstable to the RWI and exhibits vortices that eventually merge into a dominant $m=1$ mode \refe{close to the location where the lump, defined as the overdensity feature, is. In those cases as in the $q=1$ case, the lump orbits together with the $m=1$ vortex of the RWI.}
As few instabilities can produce vortices, this is a strong tie between the existence of the lump and the RWI.



\refe{\section{Discussion: \lq Lump\rq\  beyond binary black hole systems}
\label{sec:disc}}

\refe{As may be noticeable, we have not invoked any GR effect in our previous interpretation that the RWI is taking place in circumbinary discs around BBHs and is reponsible for the so-called lump feature.
More specifically, the RWI is caused by the disc sharpness sculpted by the binary torques, an effect already present qualitatively in Newtonian gravity (e.g. \citealt{artymowicz_dynamics_1994}).
Hence, it is reasonable to compare the present BBH circumbinary disc with other systems where GR is unimportant.
Therefore, in this section we discuss how the \lq horseshoe\rq\ features in protostellar and protoplanetary discs, somewhat similar to the present lump under study, could be linked to vortices and to the RWI.
We also analyse how the results of our simulations compare with numerical studies undertaken in that field.}


\refe{
As already mentioned in the Introduction, other systems than BBHs have displayed non-axisymmetric features resembling the lump: protoplanetary discs observed with ALMA exhibiting a horseshoe feature.
In that context, the RWI has been widely studied, first because it forms vortices able to trap efficiently dust particles (e.g. \citealt{meheut_dust-trapping_2012}) that can further accrete to reach planet masses, and second because the merger of those vortices may explain the horseshoe feature.
In that case, the RWI can be either triggered at the border of the dead zone (\citealt{varniere_reviving_2006}, \citealt{lyra_embryos_2008}, \citealt{lyra_planet_2009}) or at the edge of a gap cleared by a Jupiter-mass planet \citep{lyra_standing_2009}, i.e. for $q$ values much smaller than $0.1$.
It should be kept in mind that the link between the observed non-axisymmetries and the RWI is not firmly established though.
The presence of long-living vortices via the RWI has remained debated in this particular context as was questioned their survival in 3D simulations (but see \citealt{meheut_rossby_2010}), for viscous discs (with an $\alpha-$viscosity parameter - \citealt{shakura_black_1973} - larger than a given value, e.g. $10^{-4}$ in \citealt{ataiee_asymmetric_2013}, who used a planet-induced gap to trigger the RWI), or after suffering the drag force backreaction from the dusty planetary embryo (\citealt{inaba_dusty_2006}, see \citealt{lyra_embryos_2008}) - a concern that does not apply to the BBH system under study here.
Alternatively, considering a pre-existing dust trap, \cite{owen_dust_2017} claim that the accretion luminosity liberated during planet formation through pebble accretion in the trap makes the disc locally unstable to the baroclinic instability (e.g. \citealt{klahr_turbulence_2003}, \citealt{lesur_subcritical_2010}), producing vortices as well.
In summary, the ability of the RWI to produce long-living vortices in planet-forming discs, and that those are responsible for some of the non-axisymmetries of transition discs observed with ALMA, remains uncertain.
}

\refe{For the few aforementioned studies which focused on binaries (star-planet systems), the mass ratio considered is somewhat smaller than here, resulting in different configurations than the one we are focusing on (e.g. with a gap drawn by the planetary companion rather than a central cavity in the middle of a circumbinary disc).
Considering mass ratios up to $q=0.2$, \cite{ragusa_origin_2017} and \cite{calcino_signatures_2019} studied the formation of non-axisymmetries in the circumbinary disc around a star-planet system, using dusty smoothed-particle hydrodynamical simulations, and reported an azimuthal overdensity.
They do not find vortical motions in the azimuthal overdensity, which is consistent with the present study since the overdensity does not always coincide with the Rossby vortex because of the strong spiral wave forcing.
It is also worth mentioning here that we had to perform a higher-resolution run to clearly see the vortices in the simulation's outputs, especially at early times (Fig.~\ref{fig:vortex_q1}).
Nonetheless, \cite{ragusa_origin_2017} find vorticity extrema at the disc edge, but do not display the vortensity (vorticity divided by the surface density), while we find that the density decreases abruptly at the disc edge and contributes largely to the vortensity gradient (see e.g. Fig.~\ref{fig:r_rho_q}), so the absence of RWI in their simulation is not so clear.
Going to even higher mass ratios, with $q$ ranging from $0.4$ and $1$, the simulations of \cite{mosta_gas_2019} exhibit a lump and vortensity extrema at the disc edge, although the authors do not explore a relation with the RWI.
Hence, although the physical explanation behind the resemblance between the lump and the horseshoe feature remains uncertain, the present work has made a step forward by establishing a plausible link between the lump and a Rossby vortex.}
\\

\section{Conclusions}
\label{sec:ccl}

In this paper, we investigate the origin of the symmetry breaking and subsequent overdensity or azimuthal $m=1$, dubbed \lq lump\rq, in circumbinary discs around BBHs, using GRHD 2D simulations with a BBH approximate metric.
First of all, we used the symmetric mass ratio case as a singular case in which no symmetry breaking is naively expected.
We have shown that, after a few tens of orbits of the BBH, the system becomes increasingly asymmetric, resulting in a lump feature, as found in numerous studies (e.g. \citealt{shi_three-dimensional_2012}).
We further showed that this asymmetry increase is exponential.
\refe{The same conclusions are reached with a Fourier analysis: the mode $m=2$ associated with the spiral waves is forced by the binary all along the simulation, while an $m=1$ arises exponentially and eventually dominates.
Therefore, this $m=1$ asymmetry, labelled lump in the literature, is consistent with an instability.}

As a possible origin for this instability, we looked at the Rossby Wave Instability (\citealt{lovelace_negative_1978}, \citealt{lovelace_rossby_1999}), a (magneto-)hydrodynamical instability which develops at the last stable orbit of single BH accretion discs and in the dead zone of protoplanetary discs.
The RWI is characterized by an instability criterion requiring an extremum of vortensity (vorticity divided by the surface density) and by the presence of vortices.
Those vortices were invoked in the literature as a way for planetesimals to grow in mass in protoplanetary discs.
Analyzing the equal-mass case simulation, we found the RWI criterion to be fullfilled once the cavity forms, together with the presence of vortices.
Here, the vortices \refe{merge into a single vortex which allows the mass growth of the lump and orbits in Keplerian motion as well, while it is swept off by spiral waves continously.
The torques exerted by the binary cause the spiral waves and the cavity clearing as well so that it is not possible to completely disentangle the $m=2$ and $m=1$ modes in BBHs.}
Those results show that the sharp edge of circumbinary discs around BBHs is prone to the development of the RWI, which is a plausible cause for the exponential growth of the asymmetry \refe{which appears eventually as the feature dubbed lump}.

Finally, we have extended {this} study to unequal-mass BBHs {down to q=0.1}.
It had been shown by \cite{dorazio_accretion_2013} and \cite{noble_mass-ratio_2021} that the lump amplitude decreases with a decreasing mass ratio.
Here, we explain this trend by the circumbinary disc edge being less sharp as the mass ratio decreases, and therefore less prone to the development of the RWI.
\refe{Following the same principle we expect the same to be true when relaxing the circular-orbit hypothesis if it causes the edge of the disc to lose its sharpness as was shown in \cite{noble_circumbinary_2012}. Indeed, when they accounted for the BBH inspiral their disc edge became shallower and they found that the lump weakened.}
\refe{As another limitation of our work, we have considered non-eccentric binaries.
In the litterature the lump, as an overdensity feature, is less visible for eccentric binaries although the symmetry-breaking still occurs for equal-mass binaries (see e.g. \citealt{mosta_gas_2019},
\citealt{ragusa_circumbinary_2021},
\citealt{siwek_preferential_2022}). 
Nonetheless, a vortensity extremum is reported in \cite{mosta_gas_2019} for an eccentricity parameter of $0.6$, suggesting that even in this case, the RWI could be responsible for the symmetry breaking.
}

Since the RWI has been found in 2D/3D hydrodynamical/magnetic studies, for various equations of state - just like the lump feature -, if the lump is caused by the RWI, it should be a robust feature of circumbinary discs around BBHs.
Therefore, modulations of the EM flux associated to the lump are promising to allow us to distinguish BBH systems from single BHs.
\refe{Nevertheless, some parameters (e.g. the mass ratio) can affect its amplitude.}
\refe{Moreover, the thermodynamics should influence the disc sharpness, the vortex location along with it and the lump period as well, consequently.
Switching from the hydrodynamical to the magnetic case, the RWI criterion changes (see Sec.~\ref{sec:rwi}), modifying as well the location of the vortex and the lump's period.
Those aspects are beyond the scope of this paper, whose objective was to explore a possible cause for the symmetry breaking that ultimately leads to the creation of a single azimuthal overdensity known as the lump.}

As \refe{already mentioned}, the physics invoked in this paper is not specific to the BBH metric.
Neither the formation of a lump (see e.g. \citealt{shi_three-dimensional_2012}) nor the development of the RWI (\citealt{meheut_rossby_2010}) require GR effects\refe{ (see Sec.~\ref{sec:disc})}.
Very importantly for the development of the RWI, a circumbinary disc cavity is also expected to form around binary protostars (\citealt{artymowicz_dynamics_1994}, \citealt{bate_accretion_1997-1}), forming a circumbinary disc sharp edge.
As the lump or at least the asymmetry has been found in numerical studies of T-tauri star systems \citep{gunther_circumbinary_2002}, protobinary systems \citep{bonnell_formation_1994}, and, more generally, any system whose gravitational influence was treated in Newtonian gravity (e.g. \citealt{macfadyen_eccentric_2008}, \citealt{shi_three-dimensional_2012}, \citealt{dorazio_accretion_2013},\refe{\citealt{ragusa_suppression_2016}, \citealt{miranda_viscous_2017}, \citealt{calcino_signatures_2019}, \citealt{ragusa_evolution_2020}}, \citealt{tiede_how_2021}), we suggest the possibility that the present conclusions could be extended to any circumbinary disc orbiting a $q\ge0.1$ (and perhaps below $0.1$) binary system.

\section*{Acknowledgements}

RMR thanks Léna Arthur for visualization python scripts.
RMR acknowledges funding from CNES through a postdoctoral fellowship.
This work was supported by CNES, focused on Athena, the LabEx UnivEarthS, ANR-10-LABX-0023 and ANR-18-IDEX-000, and by the "Action Incitative: Ondes gravitationnelles et objets compacts" and the Conseil Scientifique de l'Observatoire de Paris.
The numerical simulations we have presented in this paper were produced on the platform DANTE (APC France) and on the HPC resources from GENCI-CINES (Grant A0100412463).

\section*{Data Availability}

The data that support the findings of this study are available from the corresponding author, R.M.R, upon request.



\bibliographystyle{mnras}
\bibliography{Zotero} 




\appendix


\bsp	
\label{lastpage}
\end{document}